\documentclass[11pt,a4paper]{article}

\usepackage{float}
\usepackage{multirow}
\usepackage{graphics}
\usepackage[cp1252,utf8]{inputenc}
\usepackage{hyperref}
\usepackage{graphicx} 
\usepackage{amsmath}
\usepackage{amsfonts}
\usepackage{amssymb}
\usepackage{bigints} 
\usepackage{relsize} 
\usepackage[top=2cm, bottom=2.5cm, left=1.5cm, right=1.5cm]{geometry}
\usepackage{multicol}
\usepackage{authblk}
\usepackage{changepage} 
\usepackage{afterpage}

\usepackage{placeins}
\usepackage[document]{ragged2e}

\usepackage{graphicx}
\usepackage{bm}
\usepackage{hyperref}
\usepackage{mathrsfs}
\usepackage{subfig}
\usepackage{caption}

\usepackage[english]{babel}

\title{\bf Shadow of charged black holes in Gauss-Bonnet gravity}
\author[a]{\bf Anish Das\thanks{anishdasslg@gmail.com,anishdas1995@bose.res.in}}
\author[b]{\bf  Ashis Saha \thanks{sahaashis0007@gmail.com, ashisphys18@klyuniv.ac.in}}
\author[a]{\bf Sunandan Gangopadhyay \thanks{sunandan.gangopadhyay@gmail.com, sunandan.gangopadhyay@bose.res.in}}

\affil[a]{\textit{Department of Theoretical Sciences, S.N.~Bose National Centre for Basic Sciences,}
	\textit{JD Block, Sector-III, Salt Lake, Kolkata 700106, India}}

\affil[b]{\textit{Department of Physics, University of Kalyani, Kalyani 741235, India}}

\date{}

\begin{document}
	\maketitle
	
	\justifying
	\begin{abstract}
		
		\noindent In this paper, we investigate the effect of higher curvature corrections from Gauss-Bonnet gravity on the shadow of  charged black holes in both $AdS$ and Minkowski spacetimes. The null geodesic equations are computed in $d=5$ spacetime dimensions by using the directions of symmetries and Hamilton-Jacobi equation. With the null geodesics in hand, we then proceed to evaluate the celestial coordinates ($\alpha, \beta$) and the radius $R_s$ of the black hole shadow and represent it graphically. The effects of charge  $Q$ of the black hole and the Gauss-Bonnet parameter $\gamma$ on the radius of the shadow $R_s$ is studied in detail. It is observed that the Gauss-Bonnet parameter $\gamma$ affects the radius of the black hole shadow $R_s$ differently for the $AdS$ black hole spacetime in comparison to the black hole spacetime which is asymptotically flat. In particular the radius of the black hole shadow increases with increase in the Gauss-Bonnet parameter in case of the $AdS$  black hole spacetime and decreases in case of the asymptotically flat black hole spacetime. We then introduce a plasma background in order to observe the change in the silhouette of the black hole shadow due to a change in the refractive index of the plasma medium. Finally, we study the effect of the Gauss-Bonnet parameter $\gamma$ on the energy emission rate of the black hole which depends on the black hole shadow radius and represent the results graphically.
		
	\end{abstract}
	
	\maketitle

	
	\section{Introduction}
	The study of black holes has been a matter of great interest ever since the existence of such objects were predicted from Einstein's general theory of relativity. The gravitational attraction of black holes is so intense that objects moving around it within a critical radius $r_c$  falls into it. This phenomena is  known as strong gravitational lensing. If the objects moving around black holes are photons coming from an illuminated  source behind the black hole, then it casts a $shadow$ in a plane which can be seen by an observer at infinity. The first study of black hole shadow was by Bardeen \cite{1}. The shadow of spherically  symmetric black holes are circular \cite{2,3}, whereas the shadows of spinning black holes are deformed \cite{4,5}. The calculation of angular radius in case of Schwarzschild black holes was carried out in \cite{1} and two observable parameters were introduced in \cite{6} which triggered the investigation of shadow of black holes in a wide variety of cases, namely, Reissner-Nordstrom, Kerr-Newman, Kerr-Sen, Kerr-Taub-NUT, Kerr-Newman NUT, non-Kerr, braneworld, regular and higher dimensional black holes \cite{7}-\cite{17}, to name a few. The Event Horizon Telescope EHT \cite{18} which is a Earth-sized millimeter-wave interferometer  spanning the earth is an ongoing project devised to gather data of the supermassive black holes of the Milky Way galaxy and our nearby galaxy. The EHT group published their recent results this year of the image of M87 \cite{19} which enabled us to see the first ever image of black hole, from which the mass and spin of the black hole were calculated.
	
	\noindent Although primarily the interest lies in studying black hole shadows, yet there has been study of shadows even for wormholes \cite{20} 
	and intense research is going on in this area in the hope of getting some possible observable predictions.\\
	The concept of working with higher curvature gravity has been of great interest recently due to its occurence in the effective low energy action of superstring theories. Now according to gauge/gravity correspondance, the higher derivative curvature terms are viewed as the corrections coming from the large N expansion of conformal field thoeries living on the boundary of the asymptotically $AdS$ spacetime in the strong coupling limit. The simplest of the higher curvature gravity is the Gauss-Bonnet (GB) gravity with  parameter $\gamma$ measuring the effect of higher curvature. The GB term in the Lagrangian is topologically invariant in $d=4$ spacetime dimensions. Hence to consider the dynamical effect of GB gravity we must work in dimensions d $ \geqslant  5$. A study of black hole shadow in higher dimension may give some insights into the nature of black holes in higher dimensions. In a recent paper \cite{21},  the structure of black hole shadows were investigated in  $(4+1)$ dimensions in the context of holography.
	Such studies indeed provide a good motivation to look at black hole shadows for Gauss-Bonnet black holes in $5$-dimensions.
	
	\noindent Shadow in case of Gauss-Bonnet gravity has been studied in \cite{22,23} considering spin and in higher dimensions. Our work investigates the shadow of charged Gauss-Bonnet gravity in $d=5$-dimensions without spin for asymptotically $AdS$ and Minkowski spaces. We also look at the black hole shadow of these black holes in a plasma medium and investigate the effects of various parameters on the shadow radius. We then proceed to investigate the energy emission rate of these black holes which depends on the radius of the black hole shadow.
	\noindent  The paper is arranged as follows. In section 1, we introduce the charged GB black hole. In section 2.1, we determine the null geodesics in $(4+1)$ dimensions and the corresponding shadow parameters  in section 3 and plot the graphs. In section 4, we study the shadow in the presence of a plasma background. In section 5, we observe how the shadow changes with the GB parameter $\gamma$ and charge $Q$. In section 6, we study the variation of energy emission rate with the frequency $\omega$. We conclude in section 7. For our calculations we consider $G$ = $c$ = $\hbar$ = 1.
	\section{Charged $AdS$ black holes in Gauss-Bonnet gravity}
	We start by writing down the Einstein-Hilbert action with a negative cosmological constant and Maxwell electrodynamics in Gauss-Bonnet gravity in $d$ spacetime dimensions. This reads \cite{24}
	\begin{equation}\label{1}
	S =\frac{1}{16 \pi G}\int{d^d x\sqrt{-g}\bigg[ \frac{(d-1)(d-2)}{l^2}+ R +\gamma\bigg(R^2 - 4R_{ab}R^{ab}+ R_{abcd}R^{abcd}\bigg) - 4\pi G F_{ab}F^{ab}\bigg] }
	\end{equation}
	\begin{equation}\nonumber
	F_{ab}=\partial_{a}A_{b}-\partial_{b}A_{a}~~
	\end{equation}
	where $\gamma \geq 0$, is the Gauss-Bonnet parameter with dimensions of $(length)^2$ and $\Lambda=-\frac{(d-1)(d-2)}{2l^2}$; $l$ denotes the $AdS$ radius, $F_{ab}$ is the Maxwell field strength tensor and  $A_{a}$ corresponds to the potential of the electromagnetic field. The field equations obtained from this action reads \cite{24} 
	\begin{eqnarray}   
	R_{ab} -\frac{1}{2}g_{ab}R &=& \frac{(d-1)(d-2)}{2l^{2}}g_{ab} +8\pi G\bigg(F_{ag}F_b^g -\frac{1}{4}g_{ab}F_{gh}F^{gh}\bigg) +\gamma\bigg[\frac{1}{2}g_{ab}\bigg(R^2-4R_{cd}R^{cd}+R_{cdef}R^{cdef}\bigg)\nonumber\\
	&-&2RR_{ab}+4R_{ac}R^{c}_{b} + 4R_{cd}R^{c~~d}_{~a~~b} -2R_{acde}R_b^{cde}\bigg].  
	\end{eqnarray}
	Assuming a spherically symmetric metric of the form 
	\begin{equation}
	ds^2 =-f(r)dt^2 +\frac{dr^2}{f(r)} + r^2h_{ij}dx^idx^j
	\end{equation}
	where $ h_{ij}dx^idx^j$ represents  the  line  element  of the $(d - 2)$-dimensional  hypersurface,  one obtains the metric coefficient $f(r)$ to be \cite{25}
	\begin{eqnarray}\label{2}
	f(r)= 1 +\frac{r^2}{2\tilde{\gamma}}\Bigg (1-\sqrt{1-\frac{4\tilde{\gamma}}{l^2}}\sqrt{1+\frac{m}{r^{d-1}}-\frac{q^2}{r^{2d-4}}}\Bigg)
	\end{eqnarray}
	where $\tilde{\gamma}=(d - 3)(d - 4)\gamma$  and  $m$, $q$  are  related  to the  gravitational  mass $M$  and  charge $Q$  as
	\begin{eqnarray}
	M=\frac{(d-2)\Sigma_k\Big(1-\frac{4\tilde{\gamma}}{l^2}\Big)}{64\pi G\tilde{\gamma}}m ~,~ \Sigma_k =\frac{2\pi^{ (\frac{d-1}{2})}}{\Gamma(\frac{d-1}{2})}
	\end{eqnarray}
	\begin{eqnarray}
	Q=\sqrt{\frac{\pi(d-2)(d-3)\Big(1-\frac{4\tilde{\gamma}}{l^2}\Big)}{2\tilde{\gamma}G}}~q~.
	\end{eqnarray} 
	\noindent In the limit $\tilde{\gamma}\to 0$, eq.(\ref{2}) simplifies to 
	
	\begin{equation}\label{3}
	f(r)= 1 + \frac{r^2}{l^2} -\frac{16\pi GM}{(d-2)\Sigma_k}\frac{1}{r^{d-3}}+\frac{GQ^2}{2\pi(d-2)(d-3)}\frac{1}{r^{2d-6}}
	\end{equation}
	which gives the Reissner-Nordstrom $AdS$ black hole solution in $d$-dimensional spacetime. From the expression of $f(r)$ in eq.(\ref{2}), we see that $\gamma$ is constrained since one must have $(1-\frac{4\tilde{\gamma}}{l^2}) \geq$ 0, which gives $\gamma \leq \frac{l^2}{4(d-3)(d-4)}$. Further, since we look for stable black holes, the stabilty condition imposes a constraint on the Gauss-Bonnet parameter, namely, $\frac{\tilde{\gamma}}{l^2 }\ge \frac{1}{36}$ \cite{24}. Hence the permissible range for $\gamma$ in $AdS_5$  black hole spacetime is
	\begin{eqnarray}\label{range}
	0.01388 \leq \gamma \leq 0.125.
	\end{eqnarray}
	
	\subsection{Geodesics in $d=5$ Gauss-Bonnet black hole}
	
	\noindent In order to find the shape of the silhouette of the shadow we first need to determine the geodesic equations traced by the photons around the black hole. To develop the formalism, we assume a test particle with rest mass $m_0$ moving around the black hole. The symmetry directions simplify the problem of finding the geodesics by determining the constants of motion associated with the direction of symmetries. The way to proceed is as follows. Let $k^{\mu}$ be the vector along the direction of symmetry and $u^{\mu}= \frac{dx^{\mu}}{d\lambda}$ be a tangent vector along a curve $x^{\mu}=x^{\mu}(\lambda)$, where $\lambda$ is the affine parameter. Then by using the Killing equation, it is easy to show that 
	\begin{equation}\label{38}
	k^{\mu}u_{\mu} = constant~
	\end{equation}
	if the trajectory $x^{\mu}$ is a geodesic \cite{26}.
	
	\noindent We once again write down the metric of the charged Gauss-Bonnet $AdS$ black hole in $d=5$ spacetime dimensions. This reads
	\begin{equation}\label{05}
	ds^2 =-f(r)dt^2 +\frac{dr^2}{f(r)} + r^2d\theta^2 + r^2\sin^2 {\theta}d\phi^2 + r^2\cos^2 \theta d\psi^2~
	\end{equation}
	where $f(r)$ in $d=5$ spacetime dimensions is given by 
	\begin{equation}\label{q}
	f(r)=1+\frac{r^2 \left(1- \sqrt{1 - \frac{8 \gamma}{l^2} + \frac{64 \gamma  M}{3 \pi  r^4}-\frac{2 \gamma  Q^2}{3 \pi   r^6}}\right)}{4 \gamma }~.
	\end{equation}
	Since the metric coefficients are time independent, therefore there is a timelike Killing vector $k^{\mu}$=(1,0,0,0,0). Eq.(\ref{38}) then gives
	\begin{eqnarray}
	k^{0}u_{0}=u_0=-E~.
	\end{eqnarray}
	The negative sign is taken for convenience. The constant $E$ can be identified as the relativistic energy per unit mass of the particle as observed by a stationary observer at infinity.
	The other symmetry directions are $\phi$ and $\psi$ since the metric coefficients are independent of these coordinates. Hence setting $k^\mu$=(0,0,0,1,0) for the $\phi$ direction and $k^\mu$=(0,0,0,0,1) for the $\psi$ direction, we obtain
	\begin{eqnarray}
	k^{3}u_{3}=u_3=L_\phi
	\end{eqnarray}
	\begin{eqnarray}
	k^{4}u_{4}=u_4=L_\psi~.
	\end{eqnarray}
	The constants $L_\phi$ and $L_\psi$ can be identified to be the angular momentum  per unit mass of the particle as seen by a stationary observer at infinity. 
	
	\noindent The geodesic equations along the directions of symmetry can now be obtained using these constants of motion. This can be done as follows. Note that 
	\begin{equation}
	u^0 = g^{0\nu}u_\nu = g^{00}u_0 = \frac{E}{f(r)}
	\end{equation}
	\begin{equation}
	u^3 = g^{3\nu}u_\nu = g^{33}u_3 = \frac{L_\phi}{r^2 \sin^2 \theta}
	\end{equation}
	\begin{equation}
	u^4 = g^{4\nu}u_\nu = g^{44}u_4 = \frac{L_\psi}{r^2 \cos^2 \theta}
	\end{equation}
	where in the last equality of the above equations we have used the contravariant components of the metric (\ref{05}).
	The above equations finally give
	\begin{eqnarray}\label{10}
	\frac{dt}{d\lambda} &=&\frac{E}{f(r)} \\
	\frac{d\phi}{d\lambda} &=&\frac{L_{\phi}}{r^{2} \sin^2\theta} \\
	\frac{d\psi}{d\lambda} &=&\frac{L_{\psi}}{r^{2} \cos^2\theta}~.
	\end{eqnarray}
	The other two geodesic equations can be derived from the Hamilton-Jacobi equation
	\begin{equation}\label{6}
	\frac{\partial S}{\partial \lambda} + \frac{1}{2} g^{\mu \sigma}\frac{\partial S}{\partial x^{\mu}}\frac{\partial S}{\partial x^{\sigma}} = 0~.
	\end{equation}
	In order to solve the Hamilton-Jacobi equation, we assume an ansatz of the form \cite{27}
	\begin{equation}\label{7}
	S=\frac{1}{2}m_{0}^2\lambda-Et +L_{\phi}\phi +L_{\psi}\psi+S_{r}(r) + S_{\theta}(\theta)
	\end{equation}
	
	\noindent where $S_{r}(r)$ and $S_{\theta}(\theta)$ are functions of $r$ and $\theta$ respectively, $\lambda$ is the affine parameter and $m_{0}$ is the rest mass of the test particle. 
	\noindent Substituting eq.(\ref{7}) in eq.(\ref{6}), we obtain
	\begin{equation}\label{qs}
	\Big(\frac{\partial S_{\theta}}{\partial \theta}\Big)^2 +   L_{\phi}^{2}\cot^{2}\theta + L_{\psi}^{2}\tan^{2}\theta + \frac{1}{2}m_0 ^2 - \frac{r^2 E^2}{f(r)} + r^2 f(r)\Big(\frac{\partial S_r}{\partial r}\Big)^2 + L_{\phi}^2 +L_{\psi}^2 =0
	\end{equation}
	which on rearranging becomes
	\begin{equation}\label{qp}
	\Big(\frac{\partial S_{\theta}}{\partial \theta}\Big)^2 +   L_{\phi}^{2}\cot^{2}\theta + L_{\psi}^{2}\tan^{2}\theta + \frac{1}{2}m_0 ^2 = \frac{r^2 E^2}{f(r)}- r^2 f(r)\Big(\frac{\partial S_r}{\partial r}\Big)^2 - L_{\phi}^2 -L_{\psi}^2 =\kappa
	\end{equation}
	where $\kappa$ is the separation constant.
	
	\noindent Now using the relation~ $p_\theta = \frac{\partial S}{\partial \theta} = \frac{\partial S_\theta}{\partial \theta}$,  we obtain
	\begin{equation}\label{1p}
	\frac{\partial S_\theta}{\partial \theta} = r^2 \frac{\partial \theta}{\partial \lambda}
	\end{equation}
	where we used the fact that $p_\theta = \frac{\partial \mathcal {L}}{\partial \dot{\theta}}$ with $\mathcal{L}$ being the Lagrangian of the particle (moving in a curved background) given by
	\begin{equation}
	\mathcal{L} = \frac{1}{2}g_{\mu \nu}\dot{x}^\mu\dot{x}^\nu~,~\dot{x}^\mu\equiv\frac{dx^\mu}{d\lambda}~.
	\end{equation}
	Similarly, using the relation~ $p_r = \frac{\partial S}{\partial r} = \frac{\partial S_r}{\partial r}$, we obtain
	\begin{equation}\label{2p}
	\frac{\partial S_r}{\partial r} = r^2 \frac{\partial r}{\partial \lambda}~.
	\end{equation} 
	Using the relations (\ref{1p}, \ref{2p}) in eq.(\ref{qp}) and setting $m_0 = 0$ to determine the null geodesics (since the rest mass of the photon is equal to zero), we obtain 
	\begin{equation}\label{8}
	r^2 \Big(\frac{d \theta}{d \lambda}\Big) = \sqrt{\Theta(\theta)}
	\end{equation}
	\noindent
	\begin{equation}\label{9}
	r^2 \Big(\frac{dr}{d \lambda}\Big)=\sqrt{R(r)}
	\end{equation}
	where\begin{equation}
	R(r)=r^4 E^2 - \Big(L^2+\kappa \Big)r^2 f(r)
	\end{equation}
	
	\begin{equation}
	\Theta(\theta)=   \kappa -   L_{\phi}^{2}\cot^{2}\theta - L_{\psi}^{2}\tan^{2}\theta
	\end{equation}
	and $\kappa$ is called the Carter constant. It is to be noted that eq.(s)(\ref{8}, \ref{9}) are the geodesic equations corresponding to $\theta$ and $r$ respectively.\\
	Eq.\eqref{9} can be cast in the following familiar form 
	\begin{equation}
	\Big(\frac{d r}{d \lambda}\Big)^2 + V_{eff}(r) = 0
	\end{equation}
	where $V_{eff}$ is the effective radial potential given by
	\begin{equation}\label{c1}
	V_{eff}(r)=\frac{f(r)}{r^2}\Big(\kappa + L^2 \Big) - E^2
	\end{equation}
	\begin{equation}\nonumber
	L^2 \equiv L_\phi ^2 + L_\psi ^2.
	\end{equation}    
	\noindent In order to find the unstable circular orbits we impose the conditions
	\begin{eqnarray}\label{60}
	V_{eff}(r)\Big\vert_{r=r_p} = 0, ~  \frac{\partial V_{eff}(r)}{\partial r}\Big\vert_{r=r_p} = 0 
	\end{eqnarray}
	and check whether $ V_{eff}(r)$ is a maxima at $r=r_p$, that is 
	\begin{eqnarray}
	\frac{\partial^2 V_{eff}(r)}{\partial r^2}\Big\vert_{r=r_p} < 0
	\end{eqnarray}
	where $r_p$ is the radius of the photon sphere.
	
	\noindent Now using eq.(\ref{c1}), the condition $V_{eff}(r=r_p) = 0$ leads to  
	\begin{eqnarray}\label{1a}
	{r_{p}}^2/f(r_p) &=& \eta + (\xi_1^2+\xi_2^2) \nonumber\\
	&\equiv& \eta + \xi^2 ~,~\xi^2 \equiv \xi_1 ^2 + \xi_2 ^2
	\end{eqnarray}
	where we have used the definitions of Chandrasekhar constants $\eta$, $\xi_1$ and $\xi_2$ \cite{27}
	\begin{eqnarray}
	\eta = \frac{\kappa}{E^2} , ~~  \xi_1 =\frac{L_{\phi}}{E},~ 
	\xi_2 = \frac{L_{\psi}}{E}~.
	\end{eqnarray}
	\noindent The boundary condition $\frac{\partial V_{eff}(r)}{\partial r}\Big\vert_{r=r_p} =0$ leads to
	\vspace{-0.5cm}

	\centering
	{\begin{equation} \label{1b}
		rf'(r)\Big\vert_{r=r_p}-2f(r=r_p)=0.  
		\end{equation}}
	\vspace{-0.5cm}
	
	\flushleft
	
	Now using $f(r)$ from eq.(\ref{q}) and it's first derivative $f'(r)$ given by
	\begin{equation}\label{70}
	f'(r) = \frac{1}{\sqrt{1 - \frac{8 \gamma}{l^2} + \frac{64 \gamma  M}{3 \pi  r^4}-\frac{2 \gamma  Q^2}{3 \pi   r^6}}}\Bigg[\frac{r}{2\gamma}\Bigg(\sqrt{1 - \frac{8 \gamma}{l^2} + \frac{64 \gamma  M}{3 \pi  r^4}-\frac{2 \gamma  Q^2}{3 \pi   r^6}}\Bigg) - \frac{r^2}{2\gamma} + \frac{4r}{l^2} - \frac{Q^2}{6\pi r^5} \Bigg]
	\end{equation}
	we have from eq.(\ref{1b}) 
	\begin{eqnarray}\label{35}
	144\pi^2 \Big(1-\frac{8\gamma}{l^2}\Big)r_{p}^8 +  \left(3072 \pi \gamma  M-4096 M^2\right)r_{p}^4 -  \left(96\pi \gamma  Q^2-384 M Q^2\right)r_{p}^2 - 9Q^2 =0~.
	\end{eqnarray}
	\flushleft
	Substituting $r_{p}^2 = x$, the above equation simplifies to a fourth order equation in $x$ given by
	\begin{eqnarray}\label{35}
	144\pi^2 \Big(1-\frac{8\gamma}{l^2}\Big)x^4 +  \left(3072 \pi \gamma  M-4096 M^2\right)x^2 -  \left(96\pi \gamma  Q^2-384 M Q^2\right)x - 9Q^2=0~.
	\end{eqnarray}
	\flushleft
	This equation can in principle be solved to obtain an exact solution. In the limit $Q\rightarrow0$ the solution reads
	\begin{eqnarray}
	x=\frac{8 \sqrt{4 l^2 M^2-3 \pi  \gamma  l^2 M}}{3 \pi  \sqrt{l^2-8 \gamma }}
	\end{eqnarray} 
	which in turn gives
	\begin{eqnarray}\label{20}
	r_p=\sqrt{\frac{8 \sqrt{4 l^2 M^2-3 \pi  \gamma  l^2 M}}{3 \pi  \sqrt{l^2-8 \gamma }}}~.
	\end{eqnarray} 
	However, for $Q \neq 0$, we solve eq.(\ref{35}) numerically.
	We set the values of mass $M$, $AdS$ radius $l$ and GB parameter $\gamma$ and then numerically solve the equation for different values of charge $Q$ to get the value of the photon sphere radius $r_p$.\\
	To find the photon sphere radius $r_p$ in asymptotically flat spacetime, we take the limit $l\to \infty$ which simplifies eq.(\ref{35}) to the form
	\begin{eqnarray}\label{402}
	144\pi^2 x^4 +  \left(3072 \pi \gamma  M-4096 M^2\right)x^2 -  \left(96\pi \gamma  Q^2-384 M Q^2\right)x - 9Q^2=0~.
	\end{eqnarray}
	Once again the solution looks simple in form when the charge $Q=0$ and reads\\
	\begin{eqnarray}\label{36}
	r_p=\sqrt{\frac{8 \sqrt{4 M^2-3 \pi  \gamma  M}}{3 \pi }}~.
	\end{eqnarray} 
	The solution in eq.(\ref{20}) reduces to eq.(\ref{36}) if we take the limit $l\to \infty$. For the $Q\neq0$ case, we once again solve eq.(\ref{402}) numerically. These results are displayed in Table 1.

	\justify
	\section{Constructing the black hole shadow}
	To obtain the shadow of the black hole, the first step is to write down the celestial coordinates which are shown in Figure 1. In $(4+1)$ dimensions, the celestial coordinates read \cite{17} 
	\begin{eqnarray}\label{clcr}
	\alpha&=&\lim_{r\to \infty}-\Big(r^2\sin\theta\frac{d\phi}{dr}+r^2\cos\theta\frac{d\psi}{dr}\Big)\nonumber \\
	\beta&=&\lim_{r\to \infty}r^2 \sin\theta\frac{d\theta}{dr} 
	\end{eqnarray}
	\begin{figure}[H]
		\centering
		\includegraphics[scale=0.45]{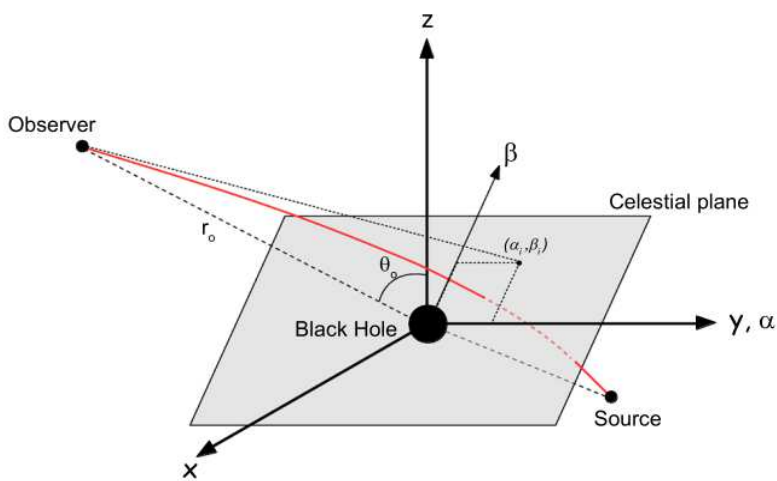}
		\caption{Basic diagram for- celestial coordinates \cite{30}}
	\end{figure}
	\noindent where $\alpha$ denotes the apparent perpendicular distance  of the shadow from the axis of symmetry (z-axis), $\beta$ denotes the apparent perpendicular distance of the shadow from it's projection on the equatorial plane, $r_0$ is the distance of the observer from the black hole and $\theta$ is the angle of inclination between the observer's line of sight and the axis of symmetry of the black hole. The celestial coordinates ($\alpha$, $\beta$) lie in a plane which passes through the black hole and is perpendicular to the line joining the black hole and the observer. This plane is known as the $celestial~ plane$.
	Using the geodesic equations given in eq.(s)(\ref{10}) and (\ref{9}), we obtain the values of $\frac{d\phi}{dr}, \frac{d\psi}{dr}$ and $\frac{d\theta}{dr}$ as given below
	\begin{eqnarray}
	\frac{d\phi}{dr} = \frac{L_\phi \csc^2 \theta}{r^2 \sqrt{E^2 - \frac{f(r)}{r^2}(\kappa + L_\phi ^2 + L_\psi ^2 )}}\\
	\frac{d\psi}{dr} = \frac{L_\phi \sec^2 \theta}{r^2 \sqrt{E^2 - \frac{f(r)}{r^2}(\kappa + L_\phi ^2 + L_\psi ^2 )}}\\ 
	\frac{d\theta}{dr} = \frac{1}{r^2}\sqrt{\frac{\kappa - L_\phi ^2 cot^2 \theta -L_\psi ^2 \tan^2 \theta}{E^2 -\frac{f(r)}{r^2}(\kappa + L_\phi ^2 + L_\psi ^2 )}}~.
	\end{eqnarray}
	Using the above relations in the expressions of celestial coordiantes ($\alpha,\beta$) and taking the limit $r\to \infty$ we get
	\begin{eqnarray}
	\alpha=-\frac{(\xi_{1}\csc{\theta} + \xi_{2}\sec{\theta})}{\sqrt{1-\frac{(\eta +\xi_{1}^2 +\xi_{2}^2)\Big(1-\sqrt{1-\frac{8{\gamma}}{l^2}}\Big)}{4\gamma}}} ~;~~  \beta=\pm\sqrt{\frac{(\eta -\xi_{1}^2\cot^2{\theta} -\xi_{2}^2\tan^2{ \theta})}{{1-\frac{((\eta +\xi_{1}^2 +\xi_{2}^2))\Big(1-\sqrt{1-\frac{8{\gamma}}{l^2}}\Big)}{4\gamma}}}}~.
	\end{eqnarray}
	Now we choose two different values of $\theta$ which are $ \theta = 0,~\frac{\pi}{2}$. When $\theta = \frac{\pi}{2}$, $L_\psi = 0$ and hence $\xi_1\equiv\xi$. On the other hand, $L_{\phi} =0$ when $\theta = 0$ which implies $\xi_2\equiv\xi$. In both cases the celestial coordinates read
	\begin{equation}
	\alpha=-\frac{\xi}{\sqrt{1-\frac{(\eta +\xi^2 )\Big(1-\sqrt{1-\frac{8{\gamma}}{l^2}}\Big)}{4\gamma}}}
	~;~~     \beta=\pm\sqrt{\frac{\eta }{{1-\frac{((\eta +\xi^2 )\Big(1-\sqrt{1-\frac{8{\gamma}}{l^2}}\Big)}{4\gamma}}}}~.
	\end{equation}
	where $\xi^2 = \xi_1 ^2 $ for $\theta = \frac{\pi}{2}$ and $\xi^2 = \xi_2 ^2 $ for $\theta = 0$.   
	\noindent Combining the coordinates $\alpha$ and $\beta$ and using eq.(\ref{1a}), we get an equation representing a circle of radius $R_s$ in the celestial plane $\alpha-\beta$, given by
	\begin{equation}\label{30}
	\alpha^2 + \beta^2 = \frac{\Big(\eta + \xi^2\Big)}{{\Big(1-\frac{(\eta +\xi^2 )\Big(1-\sqrt{1-\frac{8{\gamma}}{l^2}}\Big)}{4\gamma}}\Big)} \equiv R_s^2~.
	\end{equation}
	
	The quantity $R_s$ in eq.(\ref{30}) is the radius of the shadow given by
	\begin{eqnarray}\label{radius}
	R_s = \sqrt{\frac{\Big(\eta + \xi^2\Big)}{{1-\frac{(\eta +\xi^2 )\Big(1-\sqrt{1-\frac{8{\gamma}}{l^2}}\Big)}{4\gamma}}}}=\sqrt{\frac{\Big(\frac{r_p ^2}{f(r_p)})}{{1-\frac{(\frac{r_p ^2}{f(r_p)} )\Big(1-\sqrt{1-\frac{8{\gamma}}{l^2}}\Big)}{4\gamma}}}}
	\end{eqnarray}
	where we have used eq.(\ref{1a}) in the second equality.\\
	
	\begin{table}[ht]     
		\centering
		\hspace{-8.0cm}
		\vspace{1.8cm}
		\begin{tabular}{|c|c|c|c|}
			
			\multicolumn{4}{c}{ $AdS$ black hole ($l=1$)}\\
			\hline
			$\gamma$ & $Q$   &  $r_p$ & $R_s$ \\
			\hline
			0.04	&0&1.39975&2.00449\\
			&1&1.38123&1.98767\\
			\hline
			0.06	&0&1.47698&2.12972\\
			&1&1.4588&2.11347\\
			\hline
			0.08	&0&1.59651&2.31911\\
			&1&1.57907&2.30379\\
			\hline
			0.1&	0&1.82177& 2.66738\\
			&1&1.80591& 2.65369\\
			
			\hline
		\end{tabular}
		
		\vspace{-6.0cm}
		\hspace{9.0cm}
		\begin{tabular}{|c|c|c|c|}
			
			\multicolumn{4}{c}{Asymptotically flat black hole ($l=\infty$)
			}\\
			\hline
			$\gamma$ & $Q$   &  $r_p$ & $R_s$\\
			\hline
			0.0	&0&1.3029& 1.8426\\
			&1&1.28429& 1.82518\\
			\hline
			0.1	&0&1.21829&1.78379\\
			&1&1.19385&1.76286\\
			\hline
			
			0.2	&0&1.11106&1.71234\\
			&1&1.07415&1.68495\\
			\hline
		\end{tabular}
		\caption{\small Radius of the black hole shadow $R_s$, photon radius $r_p$ for two values of charge $Q = 0,1$ with varying values of the GB parameter $\gamma$ and $M=1$. }
	\end{table}.
	\vspace{-0.5cm}
	\vskip 0.2cm  
	\noindent In Table 1 we show the computed values of the black hole shadow radius $R_s$ and photon radius $r_p$ for different values of charge $Q$ of the black hole and GB parameter $\gamma$.
	\begin{figure}[H]
		\begin{minipage}[c]{\textwidth}
			\centering
			\subfloat[\small AdS black hole]{\includegraphics[width=0.3\textwidth]{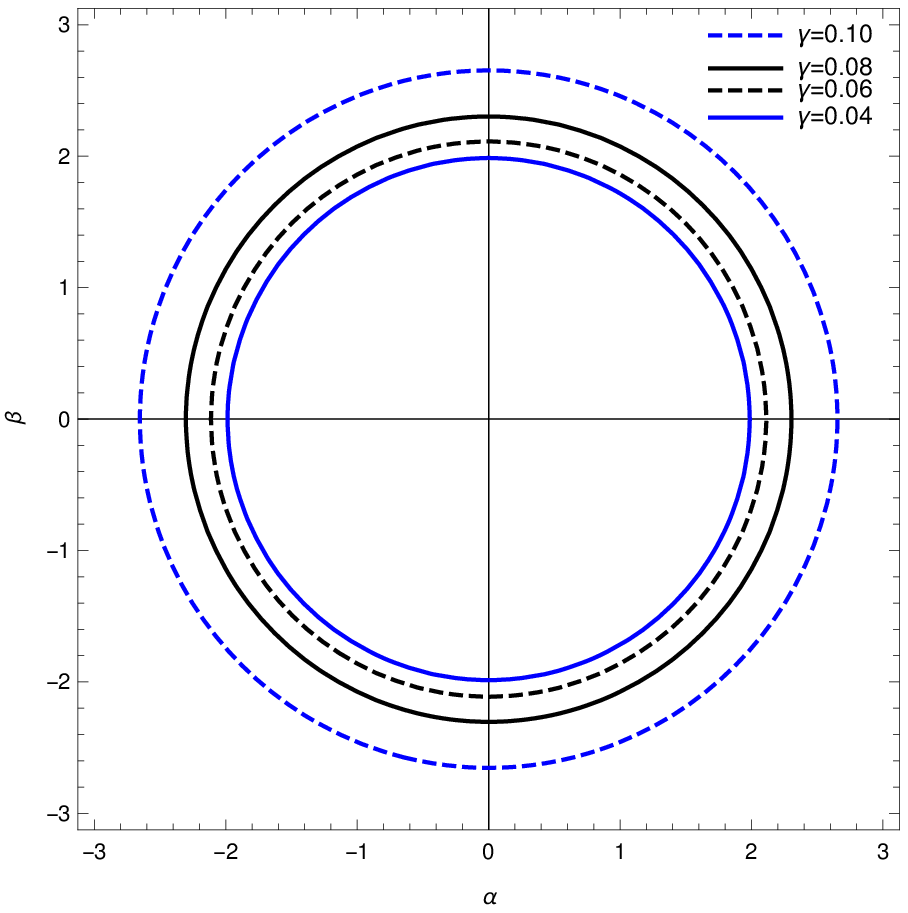}}
			\hfil
			\subfloat[\small Asymptotically flat black hole ]{\includegraphics[width=0.3\textwidth]{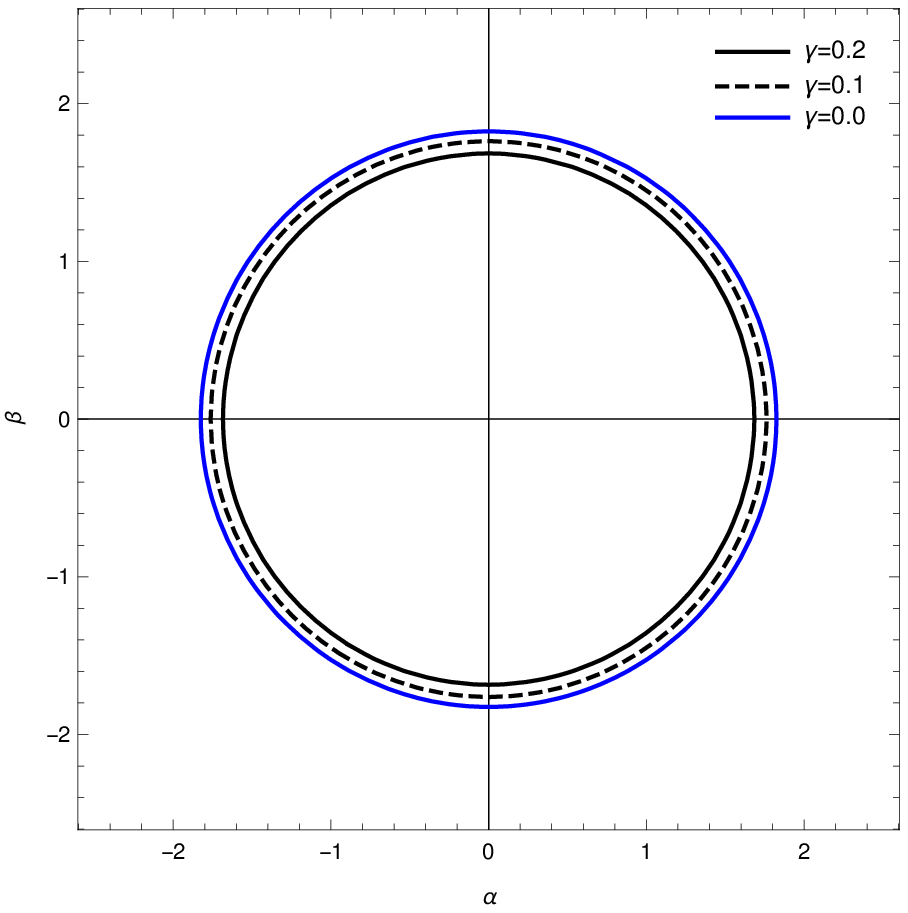}}
			\caption{\small Black hole shadow in the celestial plane ($\alpha - \beta$ plane) for various values of the GB parameter $\gamma$ with $Q=1$ along with $l=1$ for AdS black hole spacetime and $l = \infty$ for asymptotically flat black hole spacetime.}
			\label{fig:2}
		\end{minipage}
	\end{figure}  
	\begin{figure}[H]
		\begin{minipage}[t]{1.0\textwidth}
			\centering
			\subfloat[\small AdS black hole spacetime]{\includegraphics[width=0.3\textwidth]{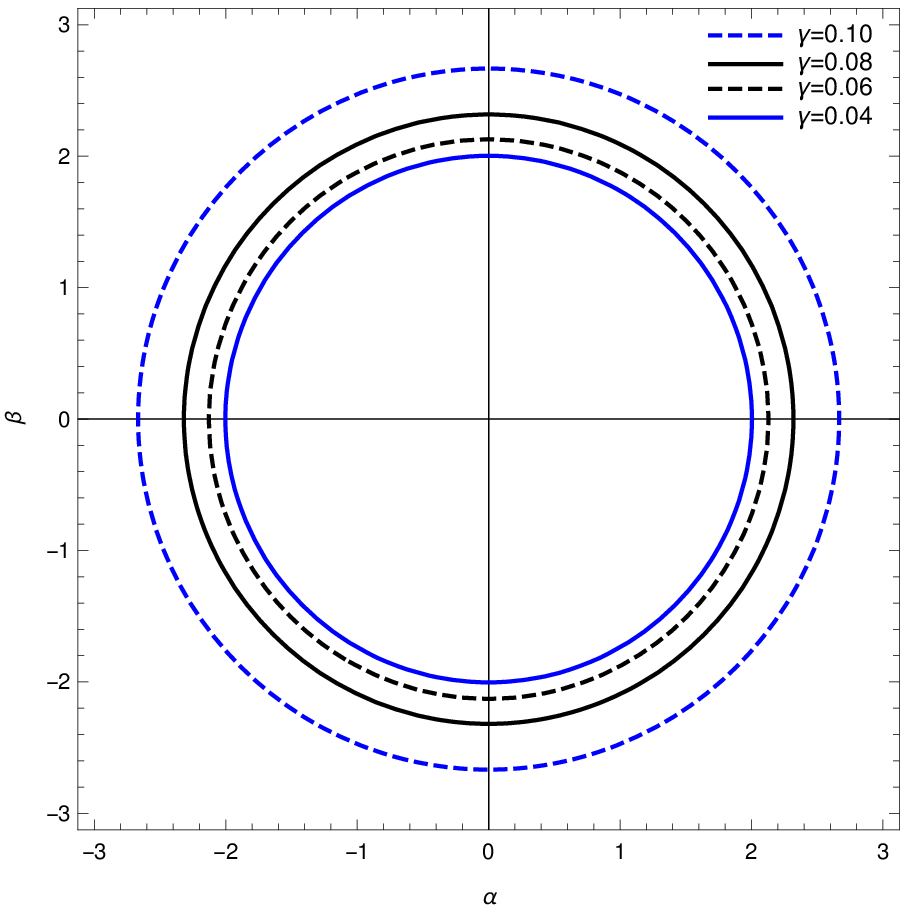}}
			\hfil
			\subfloat[\small Asymptotically flat black hole spacetime]{\includegraphics[width=0.3\textwidth]{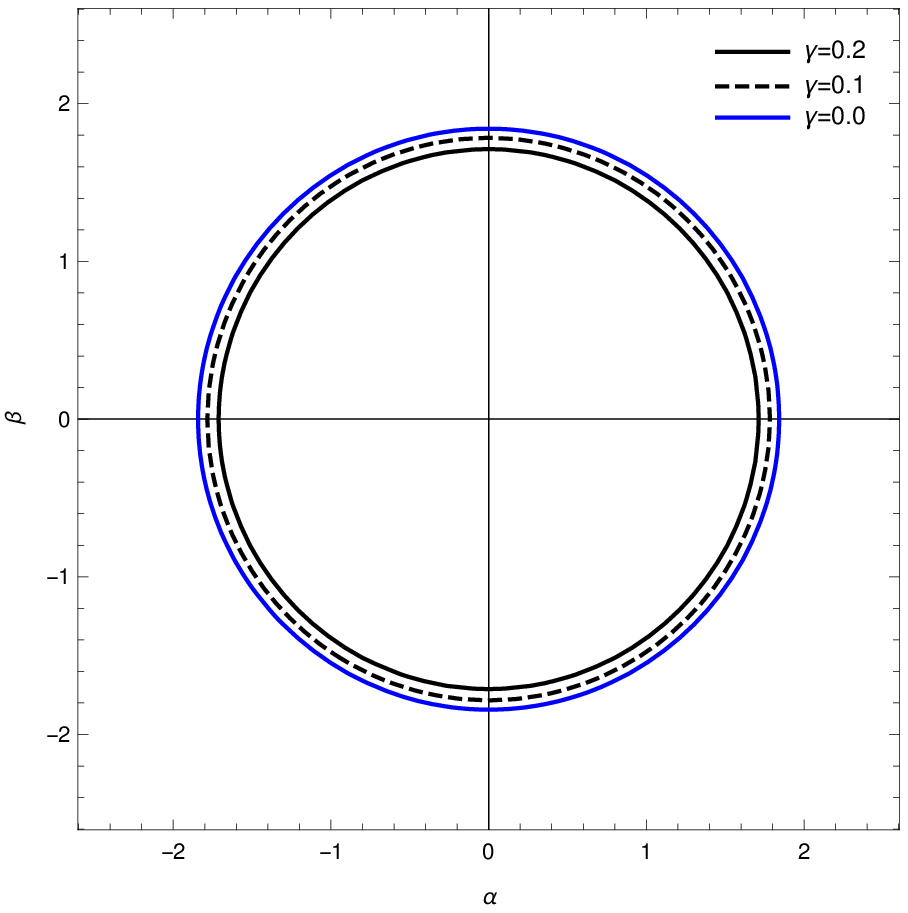}}
			\caption{\small Black hole shadow in the celestial plane ($\alpha - \beta$ plane) for various values of the GB parameter $\gamma$ with $Q=0$ along with $l=1$ for AdS black hole spacetime and $l = \infty$ for asymptotically flat spacetime.}
			\label{fig:3}
		\end{minipage}
	\end{figure}
	\noindent In Figures (\ref{fig:2}, \ref{fig:3}), the variation in the silhouette of the black hole shadow for different values of the GB parameter and charge $Q=0,1$ is shown graphically. We observe from the Figures that in case of the $AdS$ black hole spacetime ($l=1$), an increase in the value of the GB parameter $\gamma$ enlarges the size of the black hole shadow. However, on the other hand in the asymptotically flat black hole spacetime ($l= \infty$), the size of shadow reduces with an increase in the value of the GB parameter $\gamma$. We also observe that the presence of the charge $Q$ does change the radius of the black hole shadow. In fact, the size of the radius of the shadow decreases with increase in the charge of the black hole.
	
	\section{Shadow in presence of the plasma medium }
	
	\noindent In this section we shall study the effects of a plasma background on the black hole shadow. The motivation behind this approach follows from the fact that in general a black hole is surrounded by a material media which affects the geodesics of the photons passing through it. The refractive index of the plasma medium is given  by $n=n(x^i,\omega $), where $\omega$ is the photon frequency measured by an observer moving with velocity $u^{\mu}$. The plasma background modifies the Hamiltonian and introduces additional terms in the geodesics equations and thus the trajectories of the particles (in this case photons) get modified and shows a explicit frequency dependent nature. The modified effective energy of the particle in plasma medium becomes $E$ = $\hbar\omega =-p_{\alpha}u^{\alpha}$. The relationship between the $4$-momentum of the photon and the plasma frequency is given by \cite{28}
	\begin{equation}
	n^2 =1 + \frac{p_{\alpha}u^{\alpha}}{(p_{\mu}u^{\mu})^2}~.
	\end{equation}
	The refractive index $n$ is related to the plasma frequency $\omega_p$ as \cite{29}
	\begin{equation}\label{12}
	n^2 =1-\Big(\frac{\omega_p}{\omega}\Big)^2
	\end{equation}
	where  $\omega_p$ has the form
	\begin{equation}\label{13}
	\omega_p =\frac{4\pi e^2N(r)}{m_e}~.
	\end{equation}
	In the above equation $e$, $N(r)$ and $m_e$ represents the charge, number density and mass of electrons in the plasma medium respectively. As given in \cite{28,29}, the physically relevant form of $N(r)$ is assumed to be $\frac{N_0}{r^h}$. Substituting the given form of $N(r)$ in the plasma frequency $\omega_p$ and using eq.\eqref{12} we obtain the relation
	\begin{eqnarray}
	\Big(\frac{\omega_p}{\omega})^2 = \frac{k}{r^h}~,~k\geqslant 0.
	\end{eqnarray}
	The refractive index $n$ therefore takes the form 
	\begin{equation}\label{r.i.}
	n=\sqrt{1-\frac{k}{r^h}}~.
	\end{equation}
	The power $h$ characterizes different properties of the plasma medium but we shall work with $h=1$ which takes into account the minimum dependence on $r$ \cite{29,30}. Thus the expression of the refractive index with which we shall work in the rest of the paper reads
	\begin{equation}\label{r.i.}
	n=\sqrt{1-\frac{k}{r}}~.
	\end{equation} The modified form of the Hamilton-Jacobi equation in presence of the plasma medium reads \cite{28}
	\begin{eqnarray}\label{14}
	\Big(\frac{\partial S}{\partial \lambda}\Big)+\frac{1}{2} \Big[g^{\mu \sigma}\frac{\partial S}{\partial x^{\mu}}\frac{\partial S}{\partial x^{\sigma}}-(n^2 -1)\Big(\frac{\partial S}{\partial t}\sqrt{-g^{tt}}\Big)^2\Big]=0.
	\end{eqnarray}

	\subsection{Geodesics and the shadow }
	In order to investigate the effect of a plasma background we need to compute the new set of celestial coordinates. We start the analysis by computing the new set of geodesic equations by taking into account the effect of the plasma background. The set of null geodesics in presence of the plasma medium reads 
	\begin{eqnarray}\label{15}
	\frac{dt}{d\lambda} &=&\frac{n^{2}E}{f(r)} \\
	\frac{d\phi}{d\lambda} &=& \frac{L_{\phi}}{r^{2} \sin^2\theta} \\
	\frac{d\psi}{d\lambda} &=&\frac{L_{\psi}}{r^{2} \cos^2\theta}\\
	r^2\Big(\frac{ d\theta}{d \lambda}\Big) &=& \pm \sqrt{\Theta_{pl}(\theta)} \\
	r^2 \Big(\frac{d r}{ d \lambda}\Big)&=& \pm\sqrt{R_{pl}(r)}
	\end{eqnarray}
	where
	\begin{eqnarray}
	R_{pl}(r)&=&n^{2}r^{4}E^{2} - (L^{2}+\kappa)r^{2}f(r)\\
	\Theta_{pl}(\theta) &=&   \kappa -   L_{\phi}^{2}\cot^{2}\theta - L_{\psi}^{2}\tan^{2}\theta ~.
	\end{eqnarray}
	In the derivation of the geodesic equations, we follow the approach discussed in the previous section and also use the Hamilton-Jacobi equation given in eq.(\ref{14}).
	The effective radial potential in presence of the plasma background reads 
	\begin{equation}
	V^{pl}_{eff}(r)=\frac{f(r)}{r^2}\Big(\kappa + L^2 \Big) - n^{2}E^2 ~.
	\end{equation}
	The condition for the  unstable circular orbits are given by
	\begin{eqnarray}\label{91}
	V_{eff}^{pl}(r)\Bigg\vert_{r=r^{(pl)}_p}= 0~,~
	\frac{\partial V_{eff}^{pl}(r)}{\partial r}\Bigg\vert_{r=r^{(pl)}_p} = 0 
	\end{eqnarray}
	with the condition for maximizing $V_{eff}^{pl}(r)$ being given by 
	\begin{equation}
	\frac{\partial^2 V_{eff}^{pl}(r)}{\partial r^2}\Bigg\vert_{r=r^{(pl)}_p} < 0 .
	\end{equation}
	The first condition in eq.(\ref{91}) gives
	\begin{equation}\label{26}
	\eta + \xi^2 = \frac{n^2(r)r^2}{f(r)}\Bigg\vert_{r=r^{(pl)}_p}
	\end{equation} 
	and the second condition leads to
	\begin{equation}\label{25}
	\Bigg(n(r)rf'(r) - 2n(r)f(r)-2n'(r)rf(r)\Bigg)\Bigg\vert_{r=r^{(pl)}_p}=0~. 
	\end{equation}
	On replacing $f(r)$ and $f'(r)$ from eq.(s)(\ref{q}, \ref{70}) along with 
	$n'(r)=\frac{k}{2 r^2 \sqrt{1-\frac{k}{r}}}$ (obtained from eq.(\ref{r.i.})) in 
	eq.(\ref{25}), we get an equation for the radius of the photon sphere which looks too complicated and therefore we do not present it here. Further, in this case it is not possible to obtain an exact solution of eq.(\ref{25}) even in the limit $Q\to 0$. So we proceed to solve it numerically. The presence of the plasma medium introduces an extra parameter $k$ in eq.(\ref{25}). We consider two values for $k$ which are $0.2$ and $0.4$. We then obtain the values for the photon sphere radius $r_p$ by numerically solving eq.(\ref{25}). Proceeding as before, we obtain expressions for $\frac{d\phi}{dr}, \frac{d\psi}{dr}$ and $\frac{d\theta}{dr}$ which are then used to determine the celestial coordinates ($\alpha,\beta$) in presence of the plasma medium. The expressions are  
	\begin{eqnarray}
	\frac{d\phi}{dr} = \frac{L_\phi \csc^2 \theta}{r^2 \sqrt{n^2E^2 - \frac{f(r)}{r^2}(\kappa + L_\phi ^2 + L_\psi ^2 )}}\\
	\frac{d\psi}{dr} = \frac{L_\phi \sec^2 \theta}{r^2 \sqrt{n^2E^2 - \frac{f(r)}{r^2}(\kappa + L_\phi ^2 + L_\psi ^2 )}}\\ 
	\frac{d\theta}{dr} = \frac{1}{r^2}\sqrt{\frac{\kappa - L_\phi ^2 cot^2 \theta -L_\psi ^2 \tan^2 \theta}{n^2E^2 -\frac{f(r)}{r^2}(\kappa + L_\phi ^2 + L_\psi ^2 )}}~.
	\end{eqnarray}
	Using the above relations in the expressions of the celestial coordinates ($\alpha,\beta$) defined earlier, we obtain  
	\begin{eqnarray}
	\alpha =-\frac{\xi_{1}\csc{\theta} + \xi_{2}\sec{\theta}}{\sqrt{1-\frac{(\eta +\xi_{1}^2 +\xi_{2}^2)\Big(1-\sqrt{1-\frac{8{\gamma}}{l^2}}\Big)}{4\gamma}}}~ ;~
	\beta =\pm\sqrt{\frac{(\eta -\xi_{1}^2\cot^2{\theta} -\xi_{2}^2\tan^2{ \theta})}{{1-\frac{((\eta +\xi_{1}^2 +\xi_{2}^2))\Big(1-\sqrt{1-\frac{8{\gamma}}{l^2}}\Big)}{4\gamma}}}}~.
	\end{eqnarray}
	Similar to the non-plasma case, we choose two different values of $\theta$, namely, $\frac{\pi}{2}$ and $0$. In both cases, the celestial coordinates read
	\begin{equation}\label{50}
	\alpha=-\frac{\xi}{\sqrt{1-\frac{(\eta +\xi^2 )\Big(1-\sqrt{1-\frac{8{\gamma}}{l^2}}\Big)}{4\gamma}}}~;~
	\beta=\pm\sqrt{\frac{\eta }{{1-\frac{((\eta +\xi^2 )\Big(1-\sqrt{1-\frac{8{\gamma}}{l^2}}\Big)}{4\gamma}}}}~.
	\end{equation}
	By combining the celestial coordinates given in eq.(\ref{50}) and using eq.(\ref{26}), we get
	\begin{eqnarray}\label{90}
	\alpha^2 + \beta^2 = \Bigg(\frac{\Big(\frac{n^2r^2}{f(r)}\Big)}{{1-\frac{(\frac{n^2r^2}{f(r)} )\Big(1-\sqrt{1-\frac{8{\gamma}}{l^2}}\Big)}{4\gamma}}}\Bigg)\Bigg\vert_{r=r^{(pl)}_p}\equiv R_s ^2
	\end{eqnarray}
	where $R_s$ is the radius of the black hole shadow in presence of the plasma medium.\\
	
	\noindent In Tables 2, 3, we show the computed values of the black hole shadow radius $R_s$ and photon radius $r_p$ for different values of charge $Q$ of the black hole and GB parameter $\gamma$ in presence of the plasma medium.
	
	\begin{table}[H]     
		\vspace{0.25cm}
		\hspace{1.2cm}
		\begin{tabular}{|c|c|c|c|}
			
			\multicolumn{4}{c}{ k=0.2 }\\
			\hline
			$\gamma$ & $Q$   &  $r_p$ & $R_s$ \\
			\hline
			0.04	&0&1.26288&1.43158\\  			
			&1&1.24274&1.41829\\
			\hline
			0.06	&0&1.31468&1.48548\\
			&1&1.2944&1.47241\\
			\hline
			0.08	&0&1.39359&1.56276\\
			&1&1.37341&1.55104\\
			\hline
			0.1&	0&1.53788& 1.69368\\
			&1&1.51837& 1.68354\\
			
			\hline
		\end{tabular}
		\label{tab:multicol}
		\hspace{3.0cm}  		
		\begin{tabular}{|c|c|c|c|}
			
			\multicolumn{4}{c}{k=0.4}\\
			\hline
			$\gamma$ & $Q$   &  $r_p$ & $R_s$ \\
			\hline
			0.04	&0&1.12849&1.06052\\
			&1&1.10497&1.04703\\
			\hline
			0.06	&0&1.15842&1.0894\\
			&1&1.13371&1.07626\\
			\hline
			0.08	&0&1.20383&1.13058\\
			&1&1.17771&1.11801\\
			\hline
			0.1&	0&1.28534& 1.19967\\
			&1&1.25751& 1.18815\\
			
			\hline
		\end{tabular}
		\caption{Radius of the black hole shadow $R_{s}$, photon sphere radius $r_p$ with varying values of the GB parameter $\gamma$  for two values of charge $Q=0, 1$ in $AdS$ ($l=1$) black hole spacetime}
		\label{tab:multicol}
	\end{table}
	
	\vspace{-0.5cm}
	\begin{table}[H]     
		\hspace{1.2cm}
		\begin{tabular}{|c|c|c|c|}
			
			\multicolumn{4}{c}{k=0.2}\\
			\hline
			$\gamma$ & $Q$   &  $r_p$ & $R_s$ \\
			\hline
			0.0	&0&1.27491&1.69362\\
			&1&1.25564&1.67527\\
			\hline
			0.1	&0&1.18351&1.62844\\
			&1&1.15749&1.60587\\
			\hline
			0.2	&0&1.06441&1.56832\\
			&1&1.02237&1.51562\\

			\hline
		\end{tabular}
		\label{tab:multicol}
		\hspace{3.2cm}
		\begin{tabular}{|c|c|c|c|}
			
			\multicolumn{4}{c}{ k=0.4}\\
			\hline
			$\gamma$ & $Q$   &  $r_p$ & $R_s$ \\
			\hline
			0.0	& 0& 1.23863&1.52487\\
			&1&1.21834&1.50498\\
			\hline
			0.1	&0&1.13633&1.44876\\
			&1&1.1075&1.42313\\
			\hline
			0.2	&0&0.995559&1.34738\\
			&1&0.941962&1.30724\\ 			
			
			\hline
		\end{tabular}
		\caption{Radius of the black hole shadow $R_{s}$, photon sphere radius $r_p$ with varying values of the GB parameter $\gamma$ for two values of charge $Q=0,1$ in aymptotically flat ($l=\infty$) black hole spacetime}
		\label{tab:multicol}
	\end{table}
	\begin{figure}[H]
		\begin{minipage}[t]{.5\textwidth}
			\centering
			\includegraphics[width=0.6\textwidth]{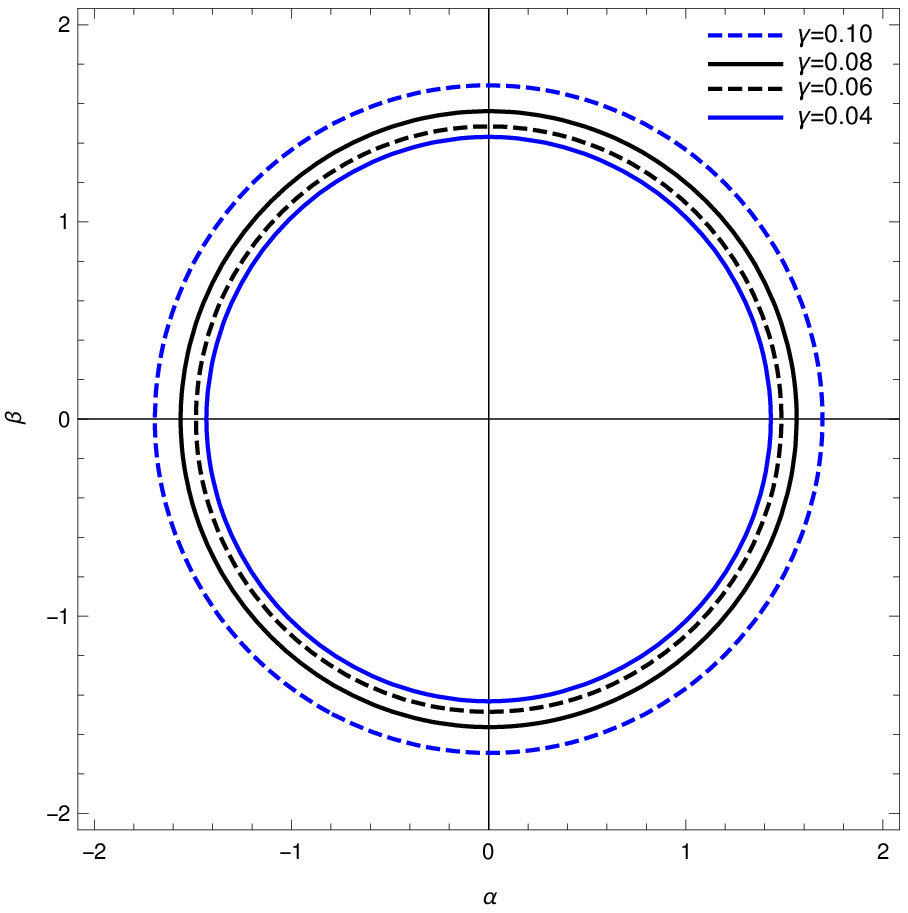}\\
			{\footnotesize $(a)~k=0.2,~Q=0$}
		\end{minipage}
		\begin{minipage}[t]{0.5\textwidth}
			\centering
			\includegraphics[width=0.6\textwidth]{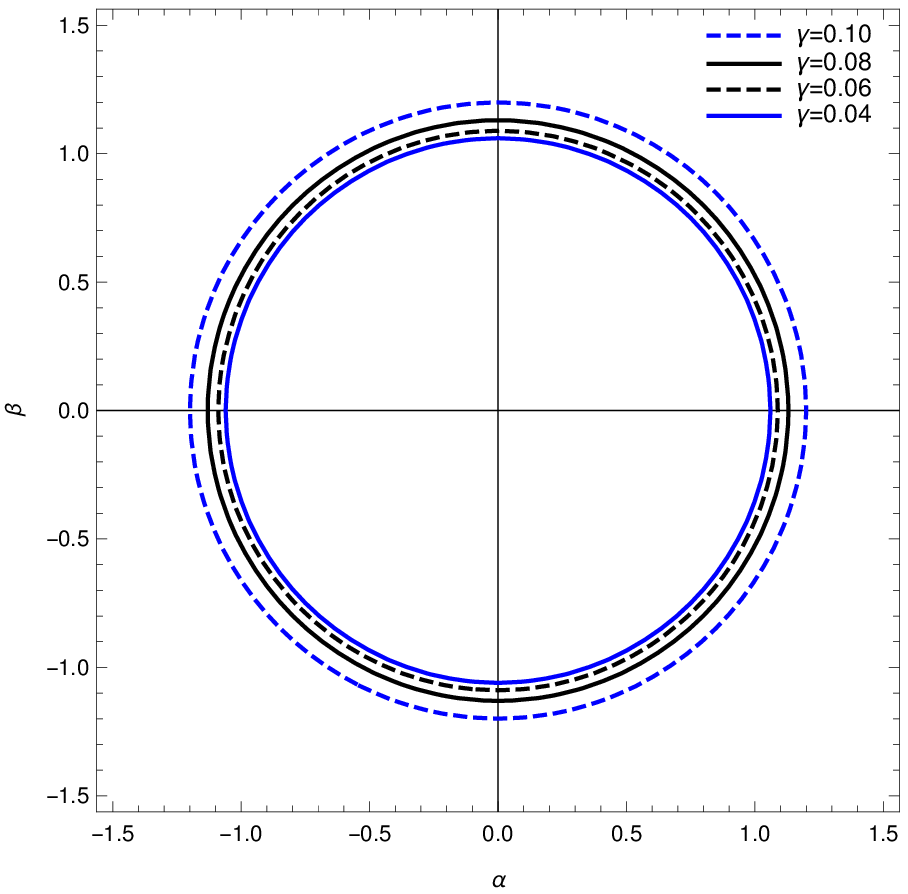}\\
			{\footnotesize $(b)~k=0.4,~Q=0$}
		\end{minipage}
	\end{figure}
	\begin{figure}[ht]
		\begin{minipage}[t]{.5\textwidth}
			\centering
			\includegraphics[width=0.6\textwidth]{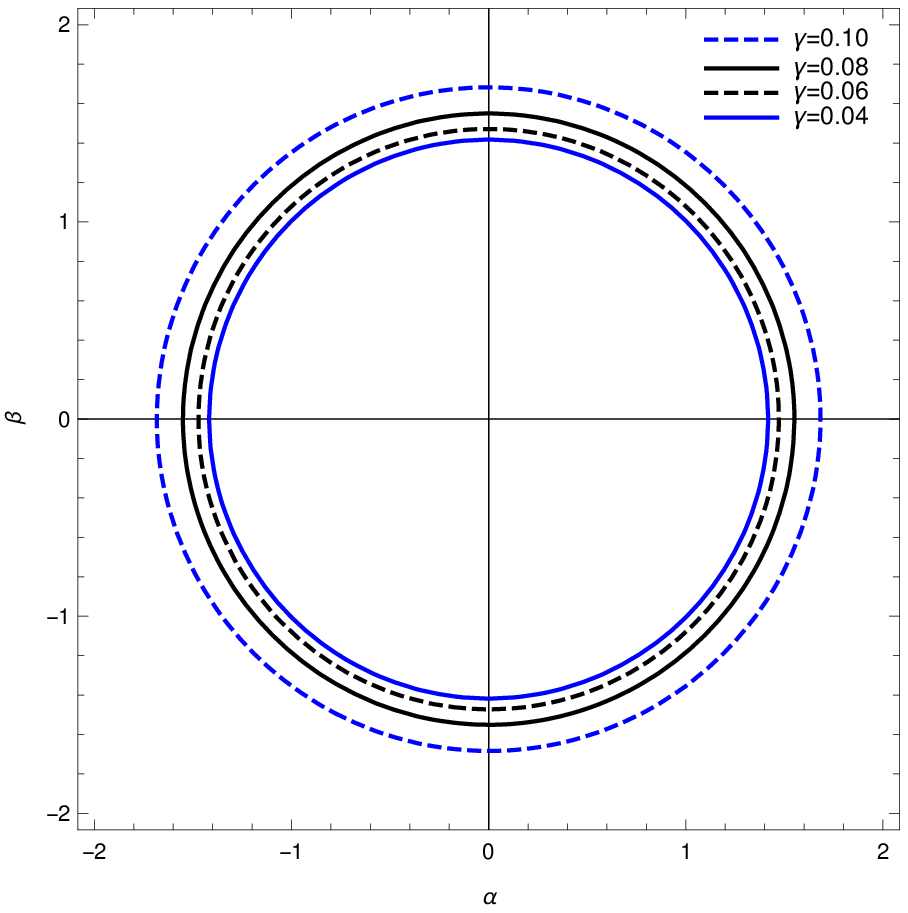}\\
			{\footnotesize $(c)~k=0.2,~Q=1$}
		\end{minipage}
		\begin{minipage}[t]{0.5\textwidth}
			\centering
			\includegraphics[width=0.6\textwidth]{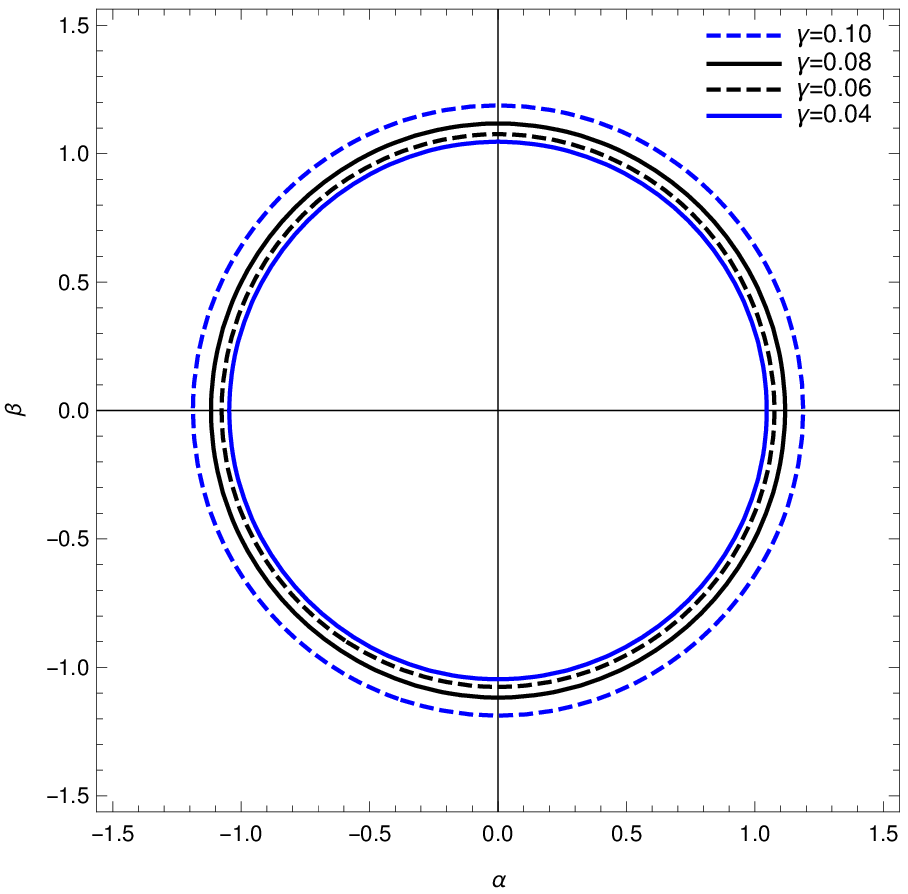}\\
			{\footnotesize $(d)~k=0.4,~Q=1$}
		\end{minipage}
		\caption{\small Black hole shadow in the celestial plane ($\alpha -\beta$) for varying $\gamma$ with charge $Q$ = 0 and 1 in $AdS$ ($l=1$) black hole spacetime for two values of plasma parameter $k$ = 0.2 and 0.4. }
		\label{fig:4}
	\end{figure}
	\noindent In Figure (\ref{fig:4}), the variation in the silhouette of the black hole shadow for different values of the GB parameter $\gamma$ and charge $Q=0, 1$ in the presence of the plasma background in $AdS$ black hole spacetime is shown graphically. We observe that with increase in $k$, the shadow radius shrinks. It is also observed that with increase in the value of the GB parameter $\gamma$, shadow size increases, and with increase in charge $Q$, the shadow radius $R_s$ decreases. 
	\begin{figure}[H]
		\begin{minipage}[t]{.5\textwidth}
			\centering
			\includegraphics[width=0.6\textwidth]{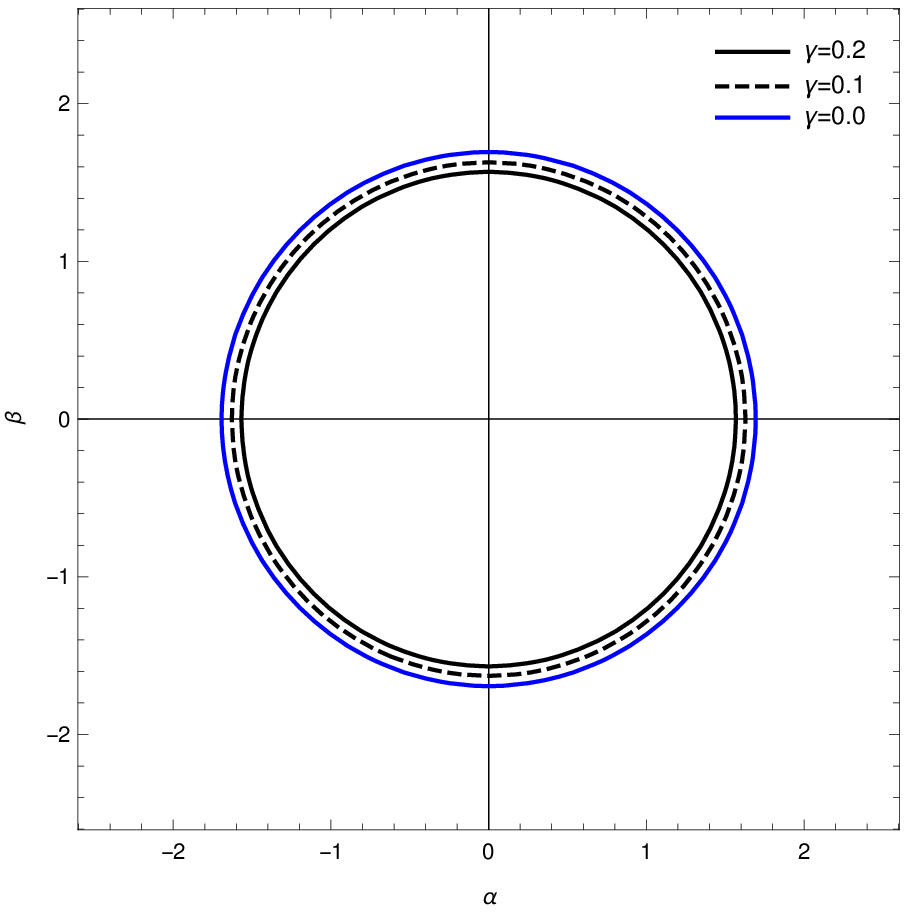}\\
			{\footnotesize $ (a)~ k=0.2,~Q=0$}
		\end{minipage}
		\begin{minipage}[t]{0.5\textwidth}
			\centering
			\includegraphics[width=0.6\textwidth]{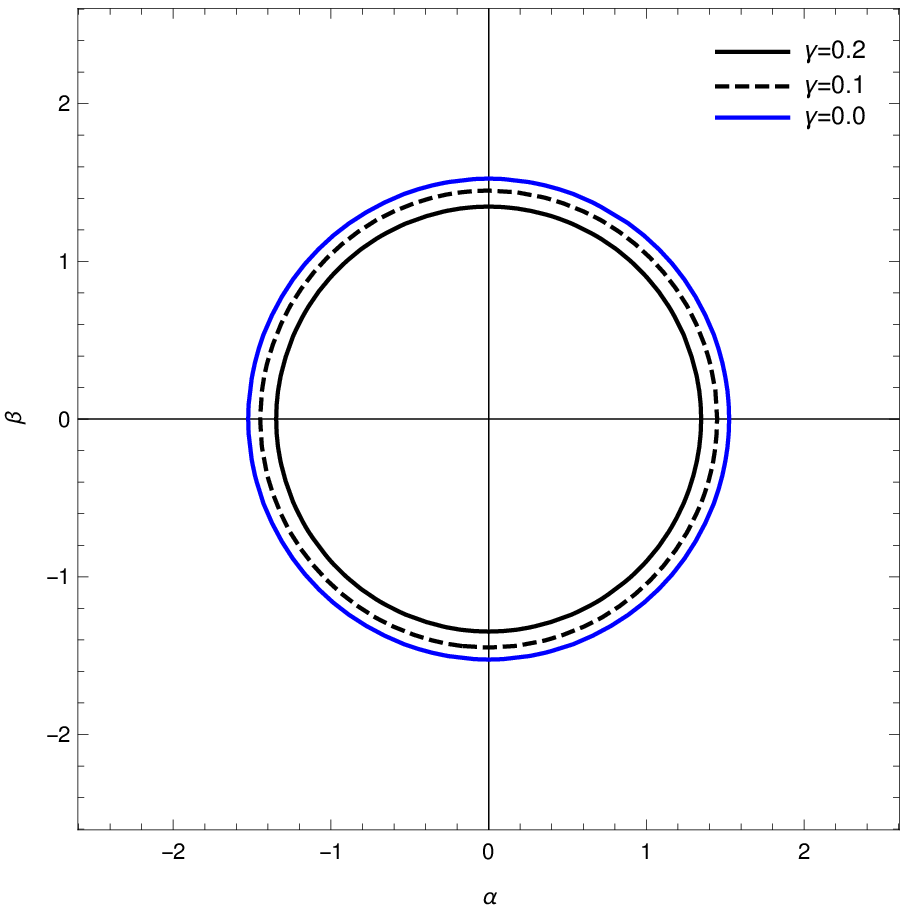}\\
			{\footnotesize $ (b)~ k=0.4,~Q=0$}
		\end{minipage}
		\vfil
		\begin{minipage}[t]{.5\textwidth}
			\centering
			\includegraphics[width=0.6\textwidth]{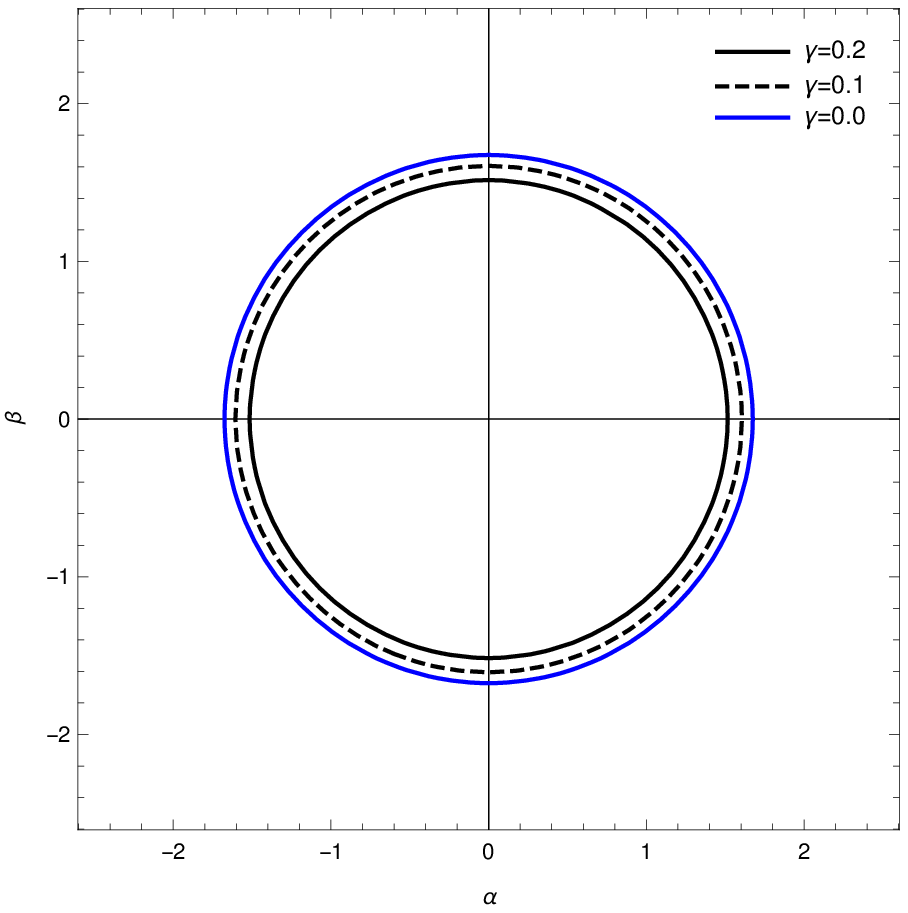}\\
			{\footnotesize $ (c)~ k=0.2,~Q=1$}
		\end{minipage}
		\begin{minipage}[t]{0.5\textwidth}
			\centering
			\includegraphics[width=0.6\textwidth]{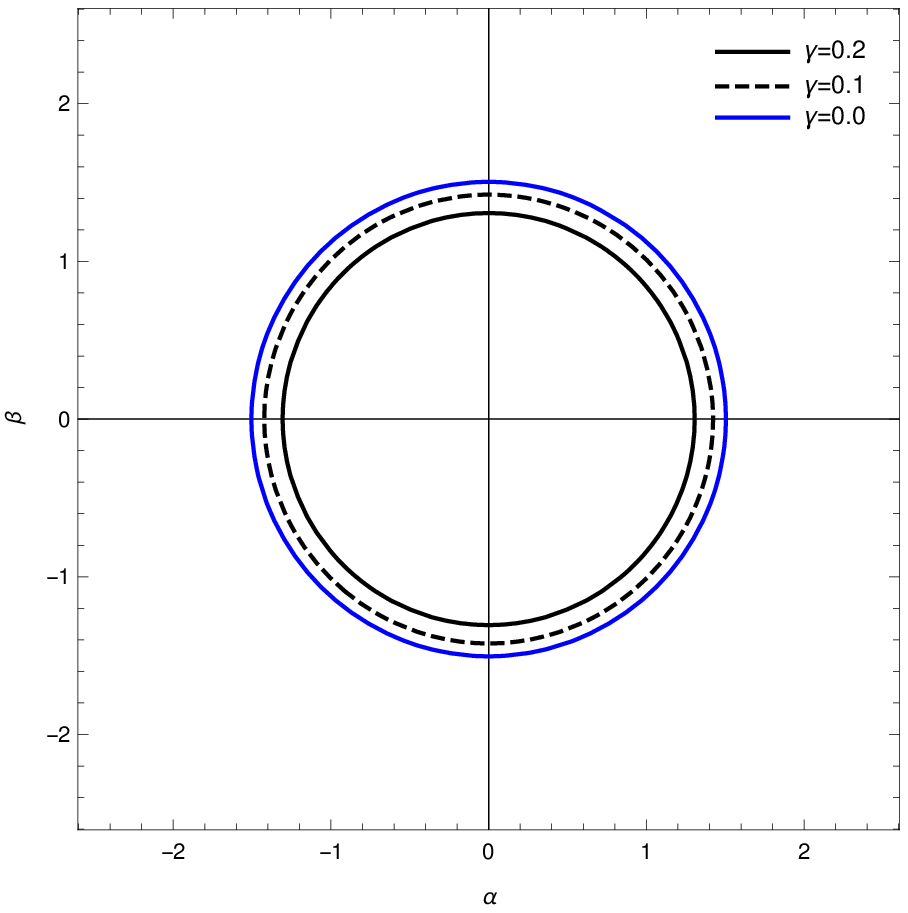}\\
			{\footnotesize $ (d)~ k=0.4,~Q=1$}
		\end{minipage}
		\caption{\small Black hole shadow in the celestial plane ($\alpha -\beta$) for varying $\gamma$ with charge $Q$ = 0 and 1 in asymptotically flat ($l=\infty$) black hole spacetime for two values of plasma parameter $k$ = 0.2 and 0.4.}
		\label{fig:5}
	\end{figure}
	
	\noindent In Figure (\ref{fig:5}), the variation in the silhouette of the black hole shadow for different values of the GB parameter $\gamma$ and charge $Q=0, 1$ in the presence of the plasma background is shown graphically for aymptotically flat ($l=\infty$) black hole spacetime. We observe that in contrast to the $AdS$ black hole spacetime, the shadow size falls with increase in the value of the GB parameter ($\gamma$). Further, we observe that with increase in $k$, the shadow size reduces. The shadow size also reduces with increase in charge $Q$.
	
	\noindent In the next section we proceed to discuss the variation of the shadow radius with the GB parameter $\gamma$ for a fixed charge $Q$ and variation of the shadow radius with charge $Q$ for a fixed GB parameter $\gamma$ for two different values of $k$.
	
	\section{Dependence of shadow radius $R_s$ on various parameters}
	We start by writing down the expression of the shadow radius $R_s$ in the presence of the plasma background. This reads
	\begin{eqnarray}
	R_s = \sqrt{\frac{\Big(\frac{n^2r_p ^2}{f(r_p)})}{{1-\frac{(\frac{n^2r_p ^2}{f(r_p)} )\Big(1-\sqrt{1-\frac{8{\gamma}}{l^2}}\Big)}{4\gamma}}}}~.
	\end{eqnarray}
	This expression of $R_s$ shows the effects of the GB parameter $\gamma$, charge $Q$ and the plasma parameter $k$ on the silhouette of the shadow in both $AdS$ ($l=1$) black hole spacetime and asymptotically flat ($l=\infty$) black hole spacetime.
	\begin{figure}[H]
		\begin{minipage}[t]{1.0\textwidth}   		   	
			\centering
			\subfloat[k=0.2,l=1]{\includegraphics[width=0.33\textwidth]{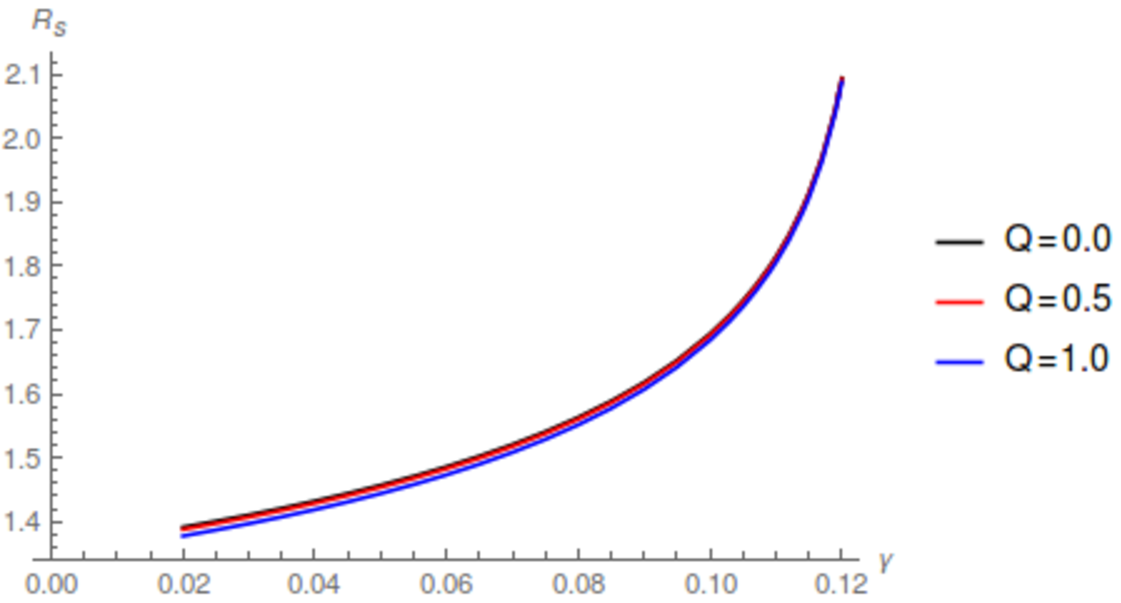}}    	
			\hfil
			\subfloat[k=0.2,l=$\infty$]{\includegraphics[width=0.33\textwidth]{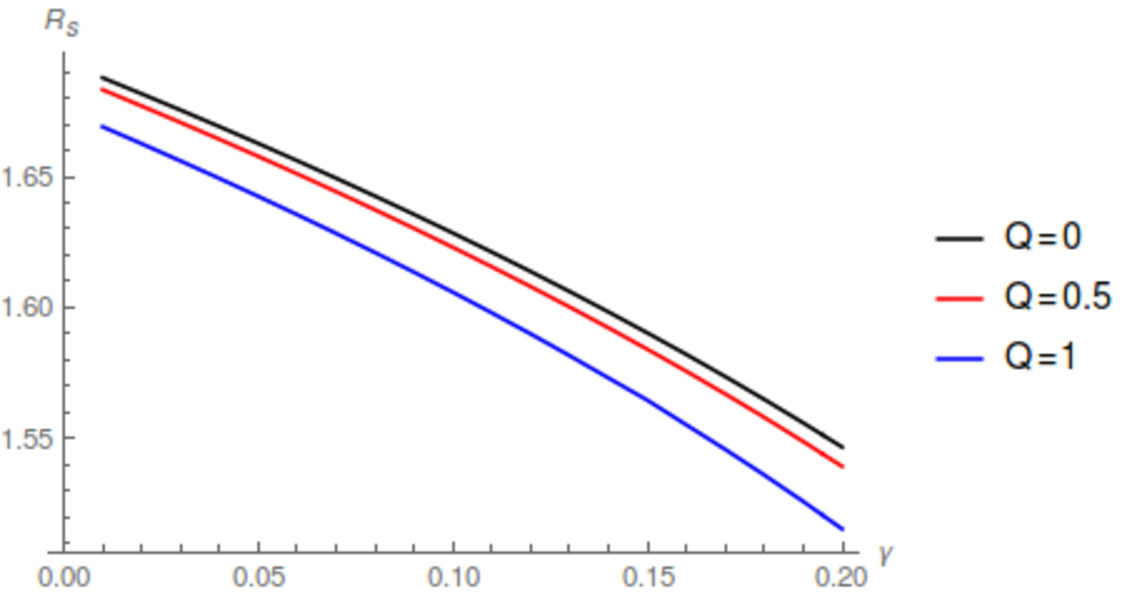}}   	
			\vfil
			\subfloat[k=0.4,l=1]{\includegraphics[width=0.33\textwidth]{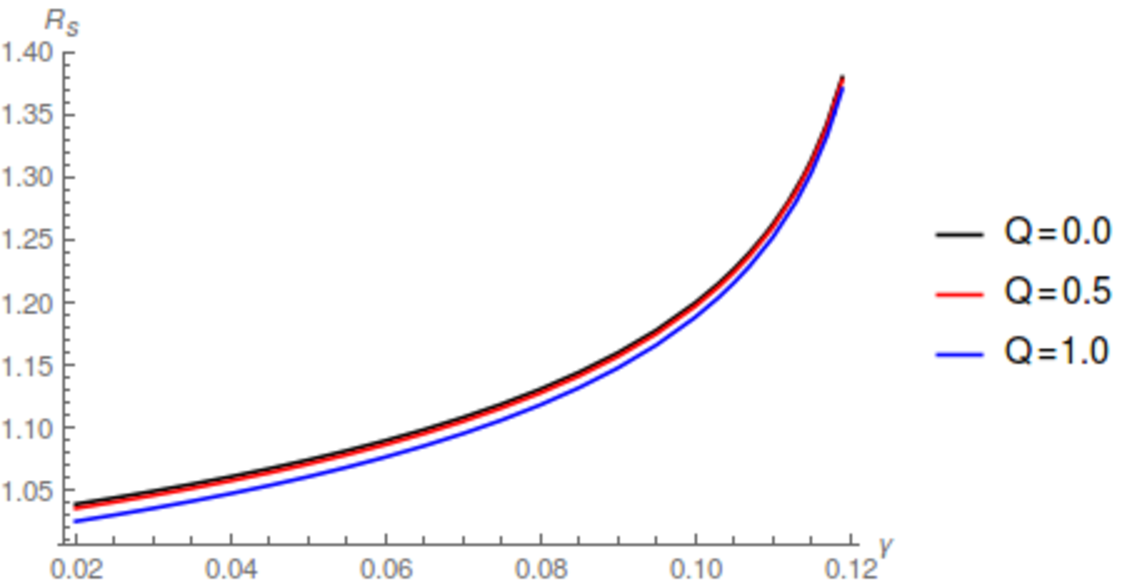}}
			\hfil
			\subfloat[k=0.4,l=$\infty$]{\includegraphics[width=0.33\textwidth]{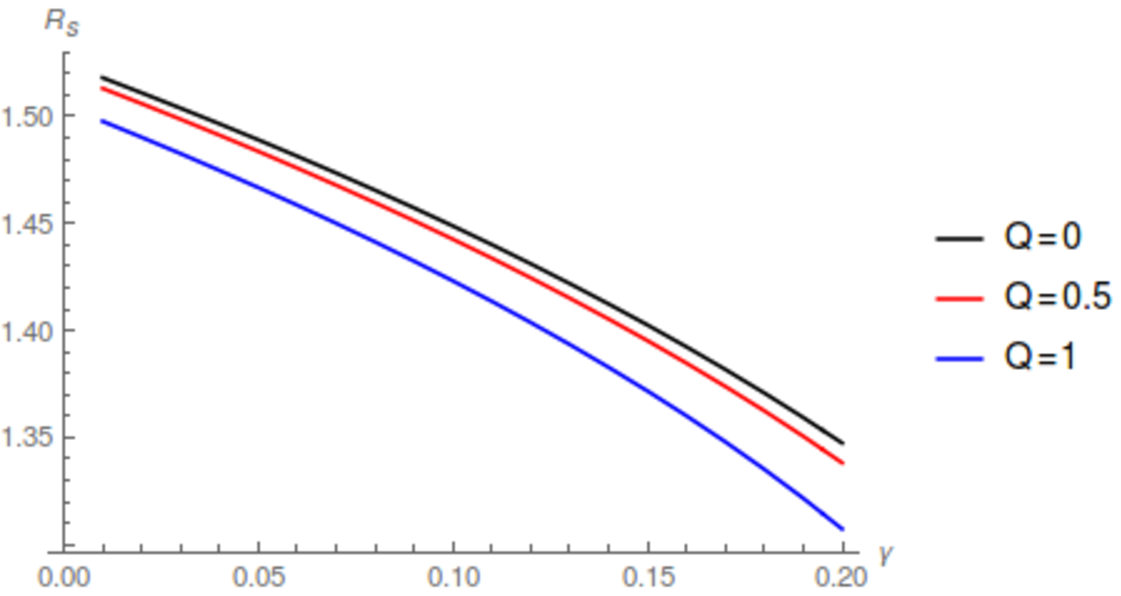}}
			\caption{\small Variation of the radius of the black hole  shadow $R_s$ with the GB parameter $\gamma$ for fixed $Q$ values in $AdS$ black hole spacetime and asymptotically flat spacetime for $k$ = 0.2, 0.4 \label{fig:11}}
		\end{minipage}
	\end{figure}
	\noindent  Figure (\ref{fig:11}) shows how the radius $R_s$ of the shadow  varies with the GB parameter $\gamma$ for fixed values of charge $Q$ in presence of the plasma medium in both $AdS$ black hole spacetime and asymptotically flat black hole spacetime. The plots are constructed for  $Q = 0,0.5$ and $1.0$. For $AdS$ ($l = 1$) black holes we found that the shadow size increase with increase in $\gamma$ for fixed $Q$, whereas the shadow size falls with increase in the value of GB parameter $\gamma$ in asymptotically flat ($l = \infty$) spacetime. Further the shadow radius $R_s$ reduces with increase in plasma parameter $k$.
	
	\begin{figure}[H]
		\centering
		\subfloat[k=0.2,l=1]{\includegraphics[width=0.33\textwidth]{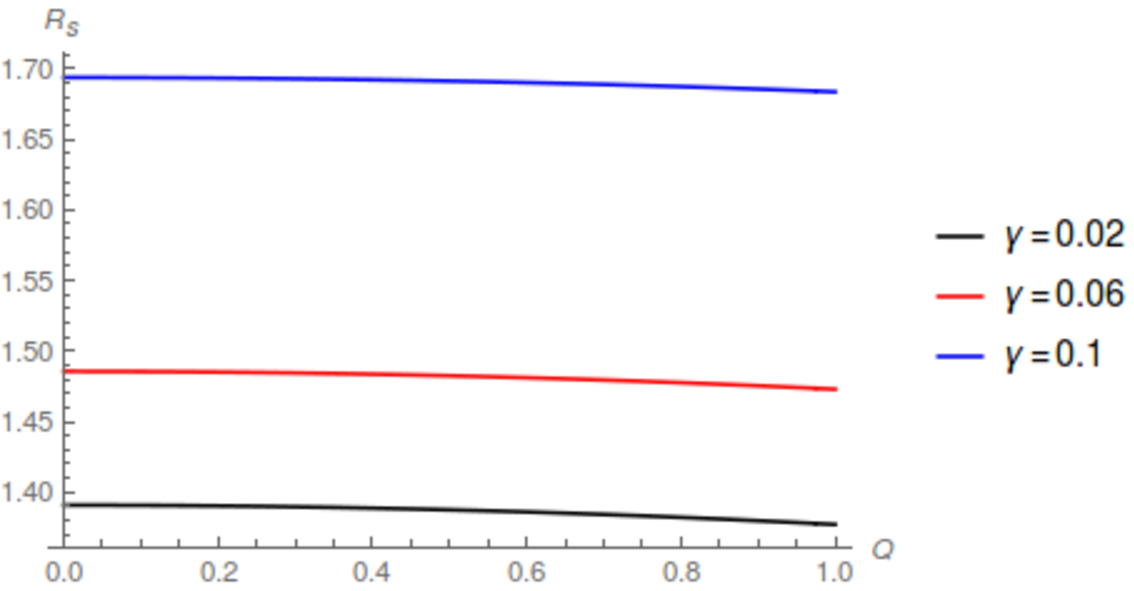}}
		\hfil
		\subfloat[k=0.2,l=$\infty$]{\includegraphics[width=0.33\textwidth]{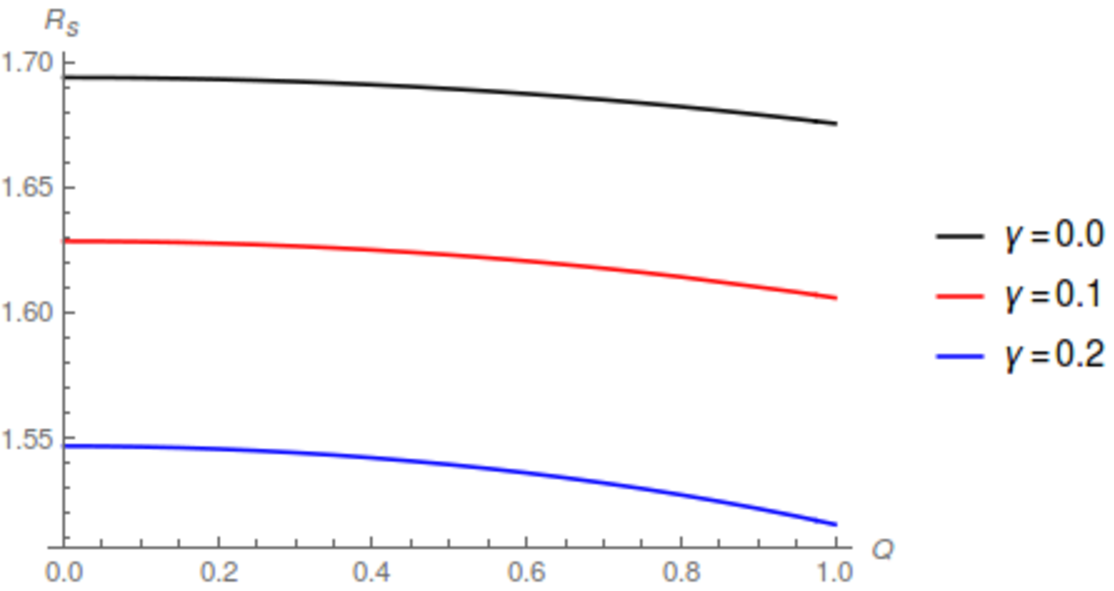}}
		
		\vfill
		\subfloat[k=0.4,l=1]{\includegraphics[width=0.33\textwidth]{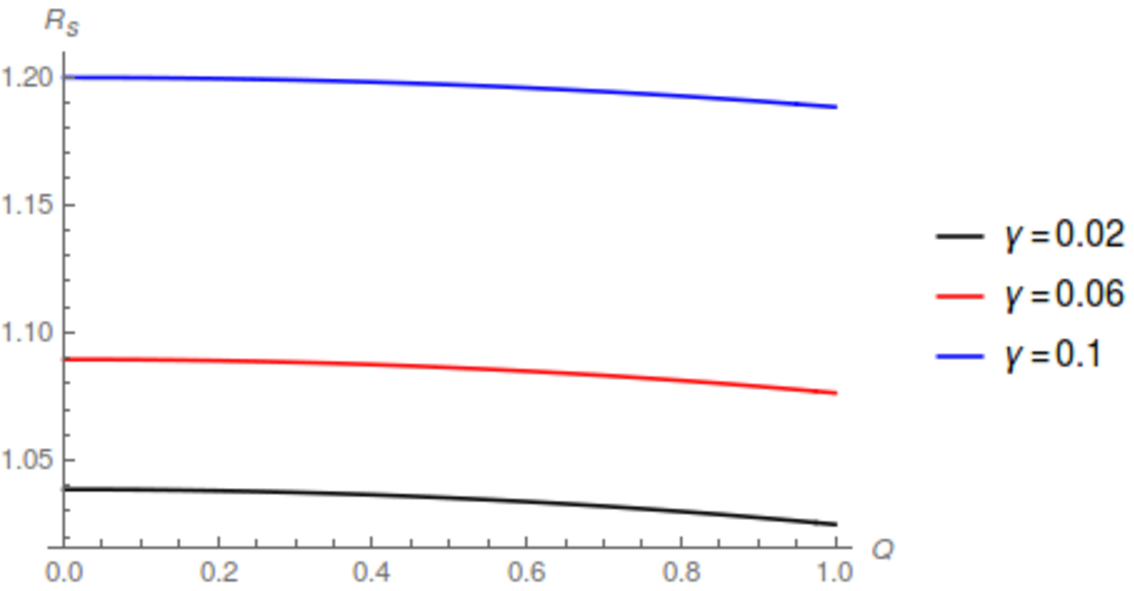}}
		\hfil
		\subfloat[k=0.4,l=$\infty$]{\includegraphics[width=0.33\textwidth]{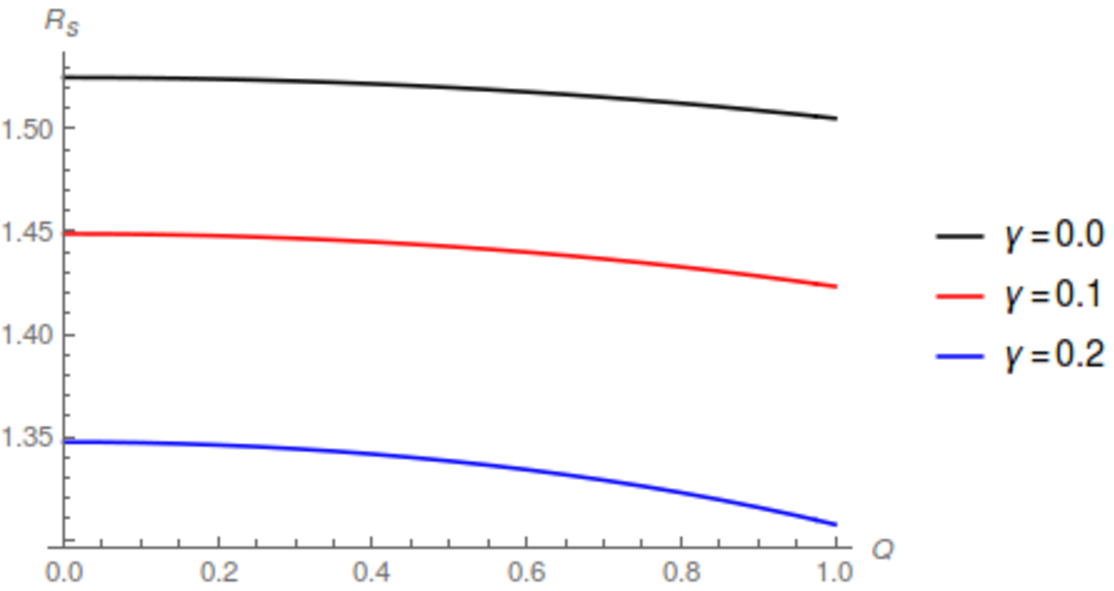}}
		\caption{\small Variation of the radius of the black hole shadow $R_s$ with $Q$ for fixed $\gamma$ values in $AdS$ black hole spacetime and asymptotically flat spacetime for $k$ = 0.2, 0.4\label{fig:10}}
	\end{figure}
	\noindent Figure(\ref{fig:10}) shows the variation of the shadow with charge $Q$ for fixed values of the GB parameter $\gamma$ in both $AdS$ ($l=1$) black hole spacetime and asymptotically flat ($l = \infty$) spacetime. It is observed that $R_s$ falls with the increase in charge $Q$. Further, $R_s$ decreases with increase in the plasma parameter.
	\section{Energy emission rate}
	
	In this section we study the energy emission rate of Gauss-Bonnet black holes in $AdS_{4+1}$ spacetime. The expression of energy emission rate reads \cite{31}
	
	\begin{equation}
	\frac{d^2 Z(\omega)}{d\omega dt}=\frac{2\pi^2 \sigma_{lim}}{exp\Big({\frac{\omega}{T_{H}}}\Big)-1}\omega^3
	\end{equation}
	where $ Z(\omega) $, $\omega$ , $T_{H}$ gives the energy, frequency and Hawking  temperature corresponding to the black hole. The Hawking temperature in $d=5$ dimensions can be obtained from eq.(\ref{2}) to be
	\begin{equation}\label{003}
	T_H = \frac{f'(r)}{4 \pi}\Bigg\vert_{r=r_+} = \frac{(4Q^2 r_{+}-128Mr_{+} ^3)+(256Mr_{+} ^2-12Q^2)}{96\pi^2 r_{+}^7}    
	\end{equation}
	where $r_{+}$ is the radius of the event horizon of the black hole.
	
	\noindent The  expression  for $\sigma_{lim}$, which is  the  limiting  constant  value is expressed in $d$ spacetime dimensions as \cite{32,33}
	\begin{equation}
	\sigma_{lim}=\frac{\pi^{\frac{d-2}{2}}R_{s}^{d-2}}{\Gamma \left( \frac{d}{2} \right)}
	\end{equation}
	where $R_s$ is the radius of the shadow. In $d=5$ dimensions, $\sigma_{lim}$ reads
	\begin{equation}
	\sigma_{lim}\approx\frac{4\pi R_s^3}{3}~.
	\end{equation}
	\begin{figure}[H]
		\centering
		\subfloat[k=0,Q=0]{\includegraphics[width=0.4\textwidth]{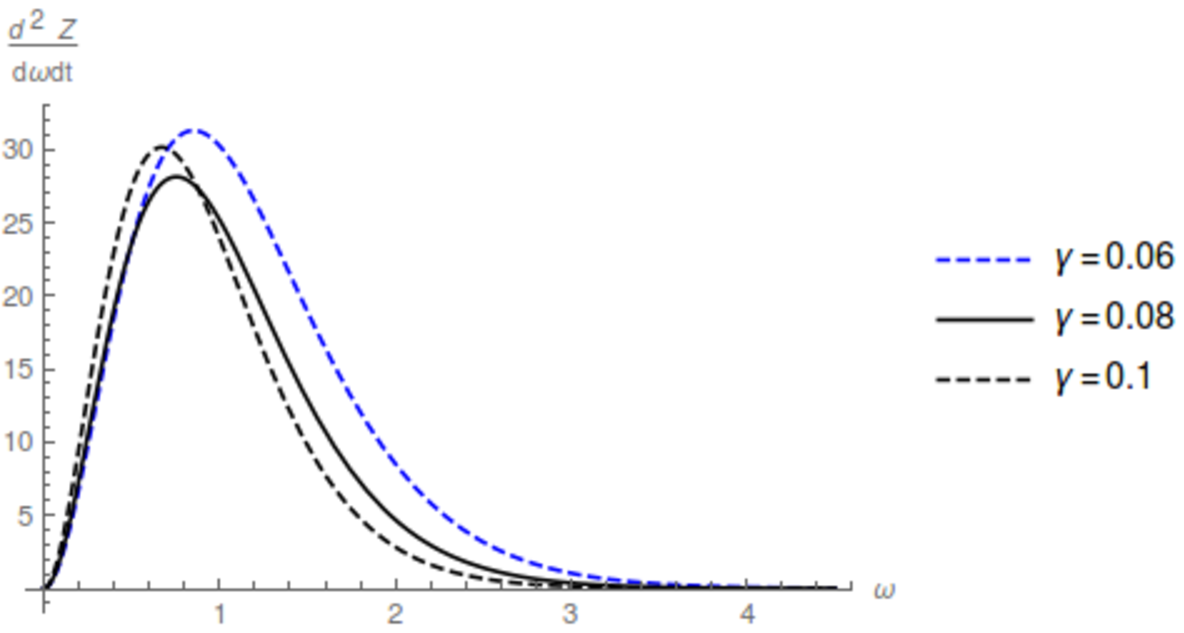}}
		\hfil
		\subfloat[k=0,Q=1]{\includegraphics[width=0.4\textwidth]{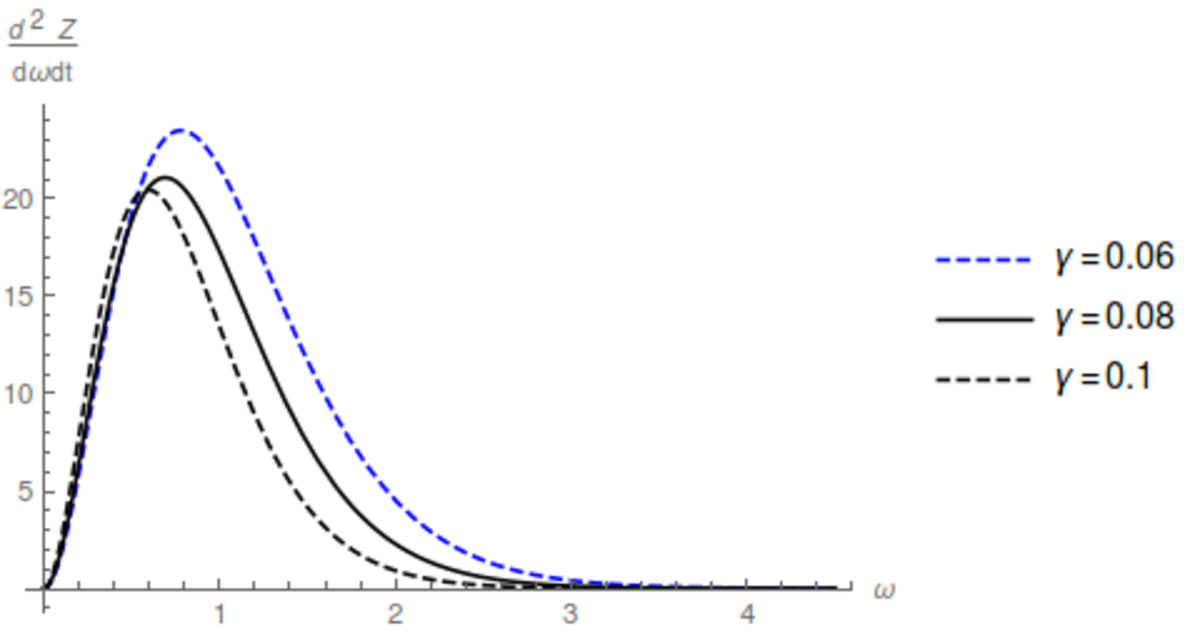}}
		
		\vfill
		\subfloat[k=0.2,Q=0]{\includegraphics[width=0.4\textwidth]{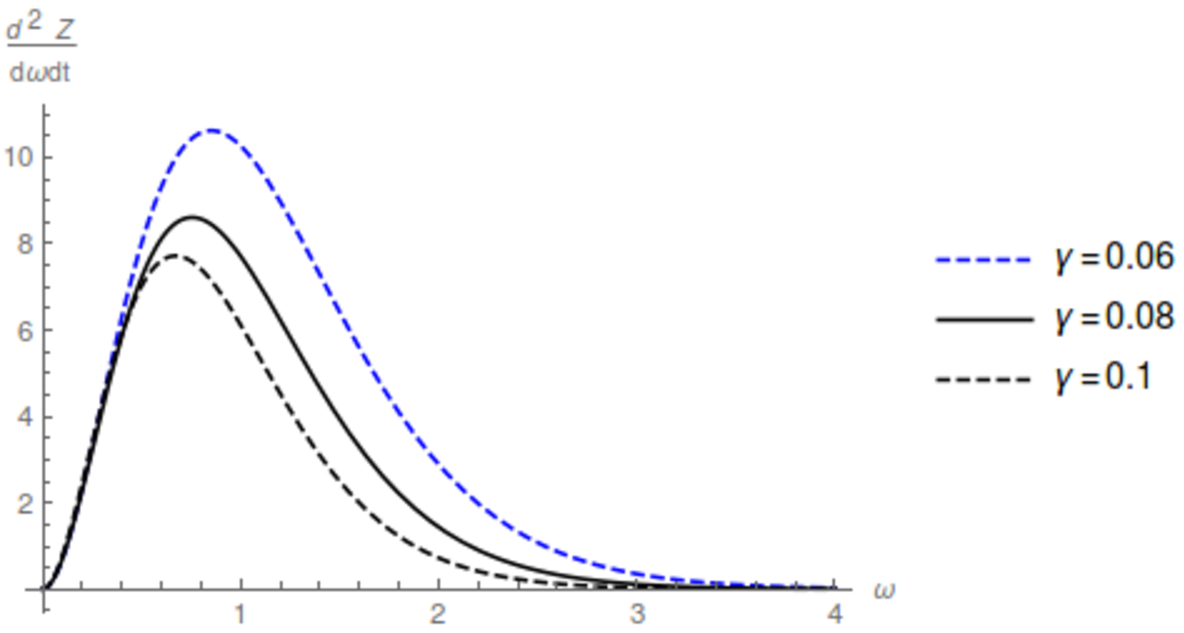}}
		\hfil
		\subfloat[k=0.2,Q=1]{\includegraphics[width=0.4\textwidth]{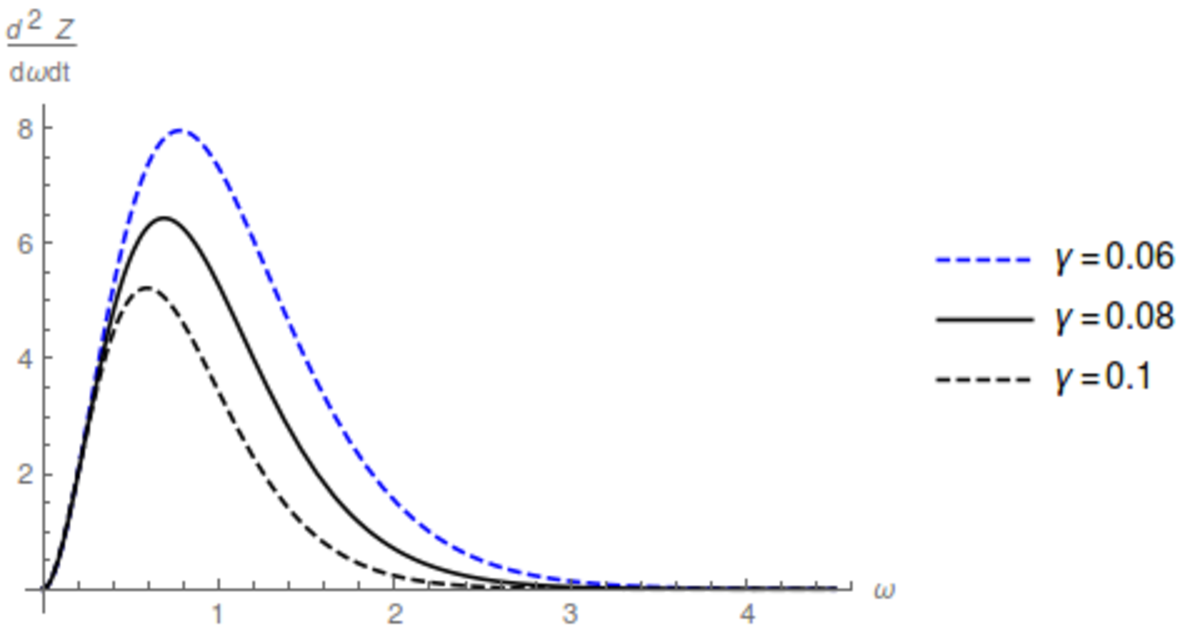}}
		\caption{\small Variation of energy emission rate ($\frac{d^2 Z}{d\omega dt}$) with frequency ($\omega$) for fixed values of Gauss Bonnet parameter $\gamma$ with $Q = 0,1$ and $k=0.0,0.2$ in $AdS$ ($l=1$) black hole spacetime}.
		\label{fig:1}
	\end{figure}       
	\begin{figure}[H]
		\centering
		\subfloat[k=0,Q=0]{\includegraphics[width=0.4\textwidth]{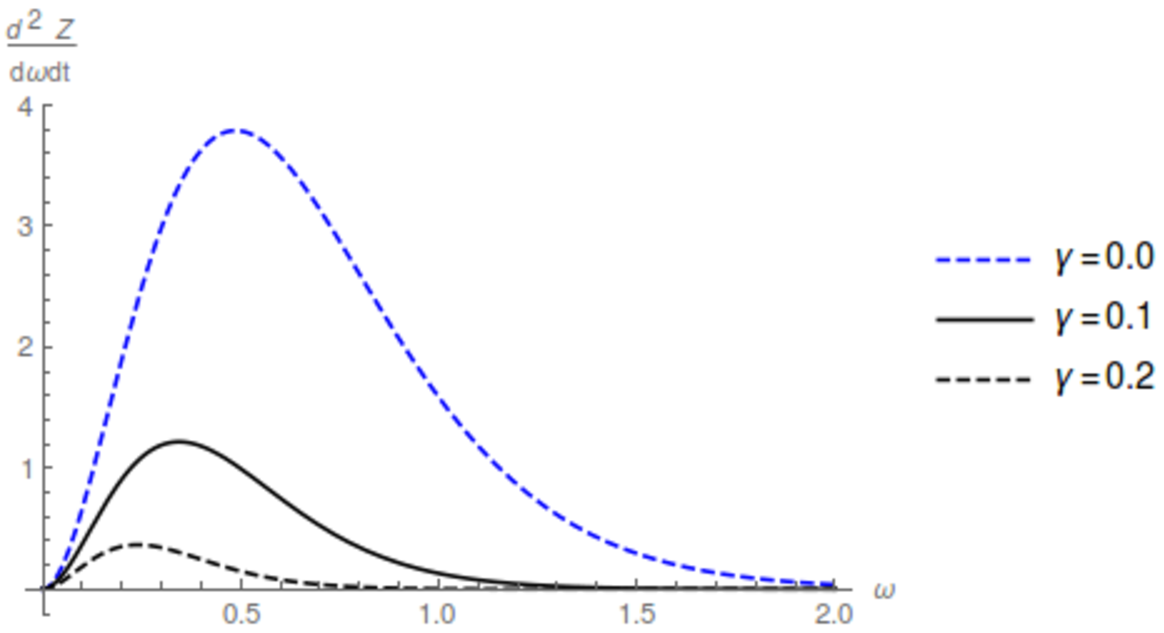}}
		\hfil
		\subfloat[k=0,Q=1]{\includegraphics[width=0.4\textwidth]{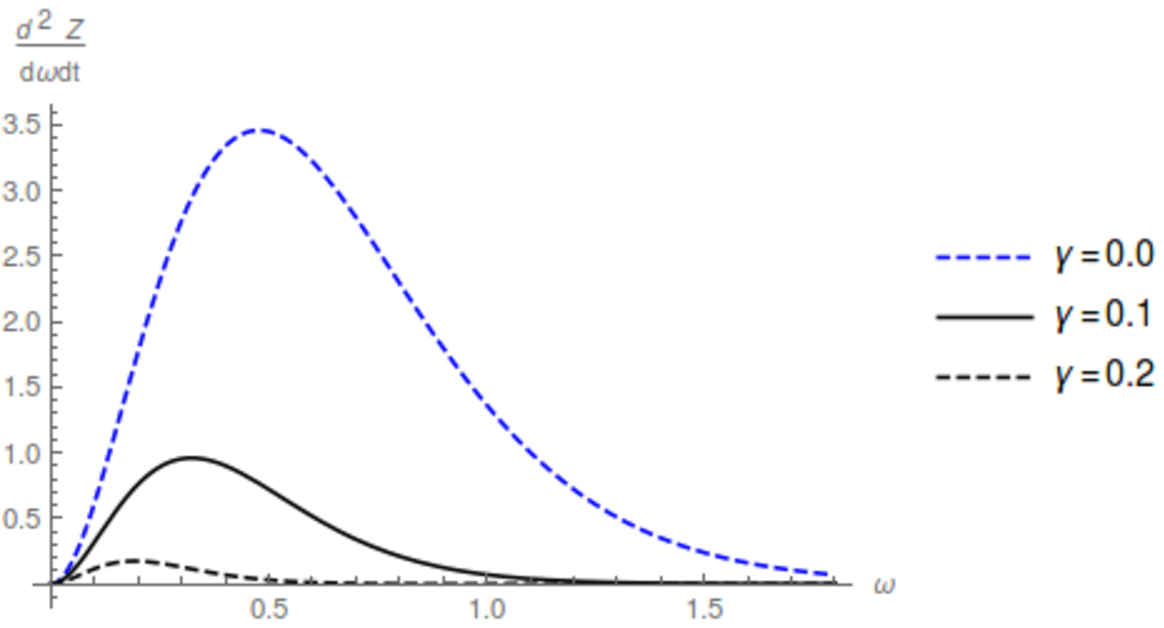}}
		
		\vfill
		\subfloat[k=0.2,Q=0]{\includegraphics[width=0.4\textwidth]{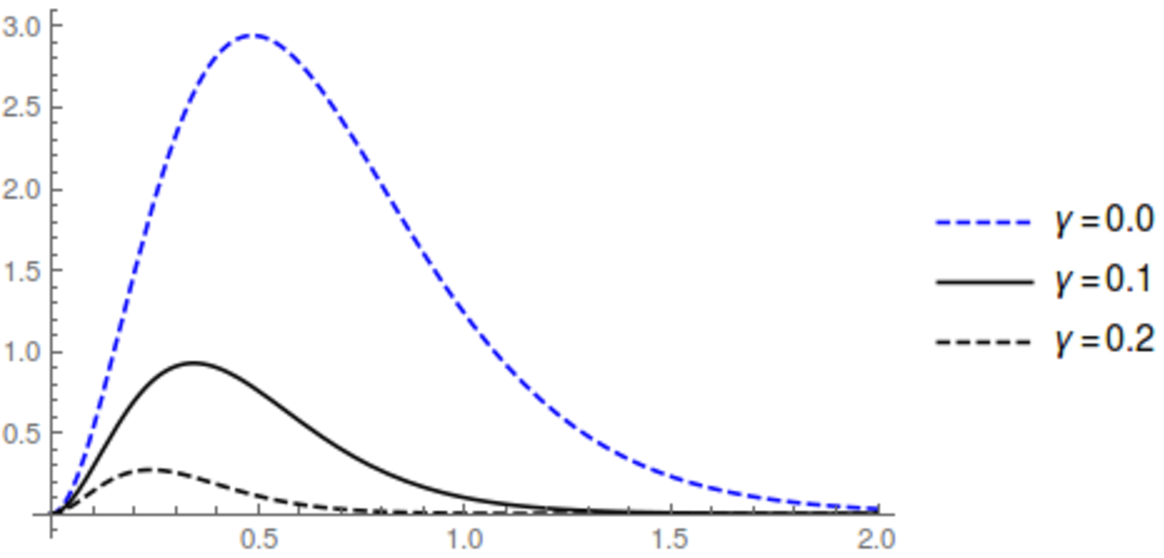}}
		\hfil
		\subfloat[k=0.2,Q=1]{\includegraphics[width=0.4\textwidth]{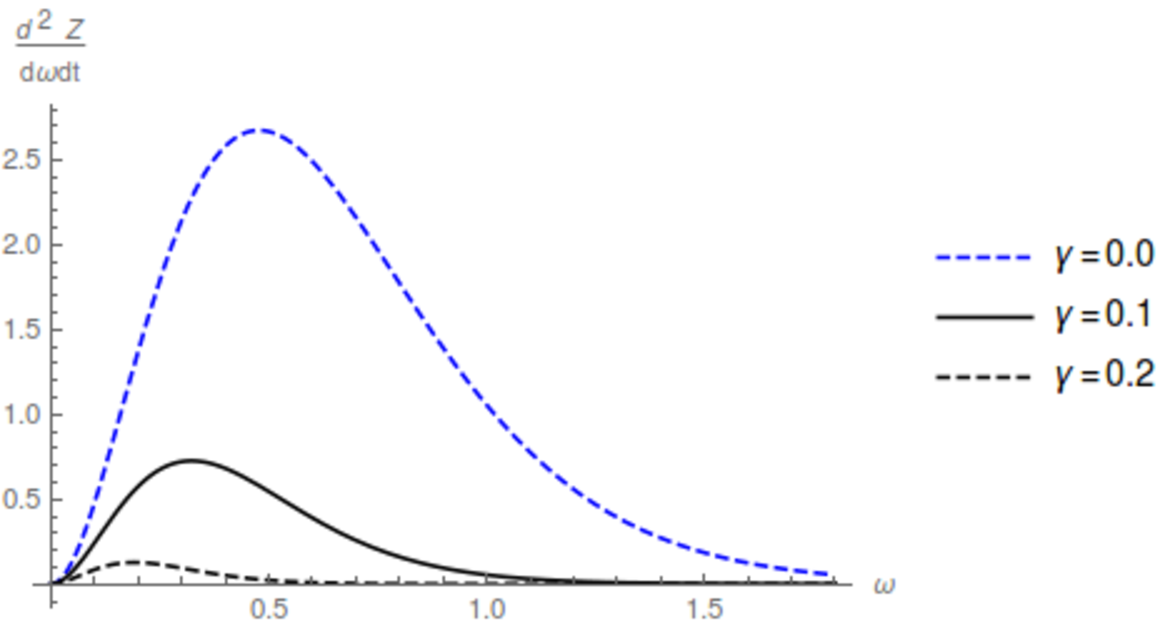}}
		\caption{\small Variation of energy emission rate ($\frac{d^2 Z}{d\omega dt}$) with frequency ($\omega$) for fixed values of Gauss Bonnet parameter $\gamma$ with $Q = 0,1$ and $k=0.0,0.2$ in asymptotically flat ($l=\infty$) black hole spacetime }
		\label{fig:7}
	\end{figure}  
	\noindent The form of the energy emission rate in $d=5$ dimensions therefore becomes
	\begin{equation}
	\frac{d^2 Z(\omega)}{d\omega dt}=\frac{8\pi^{3}R_{s}^{3}}{3\Big(exp\Big({\frac{\omega}{T_{H}}}\Big)-1\Big)}\omega^3 ~. 
	\end{equation}
	Figures (\ref{fig:1}, \ref{fig:7}) show the variation of energy emission rate $\frac{d^2 Z(\omega)}{d\omega dt}$ with frequency $\omega$ for fixed vaues of $\gamma$ with charge $Q = 0,1$ and $k = 0,0.2$. The plots are shown for both $AdS$ ($l=1$) black hole spacetime and asymptotically flat ($l=\infty$) black hole spacetime. We observe that the energy emission rate decreases with increase in $\gamma$ for both $AdS$ black hole spacetime and aymptotically flat spacetime. The presence of the plasma medium reduces the energy emission rate drastically  in $AdS$ black hole spacetime. The effect is comparatively less in case of aymptotically flat spacetime.      
	\section{Conclusion}
	We now summarize our findings. In this paper we investigate the shadow of charged Gauss-Bonnet black holes for an infinitely distant observer in $d=5$ spacetime dimensions. We first compute the null geodesic equations in $d=5$ dimensions for charged black hole in Gauss-Bonnet gravity in both aymptotically $AdS$ and Minkowski spacetimes. We then obtain the celestial coordinates ($\alpha,\beta$) by using the null geodesics which in turn gives the radius of the shadow $R_s$. We compute the values of photon radius $r_p$ and shadow radius $R_s$ taking into consideration the effect of various parameters and represent them graphically. We observe the effect of the Gauss Bonnet parameter $\gamma$ on $R_s$ which yields contrasting results for $AdS$ black hole and asymptotically flat black hole spacetime. We infer from the plots that for $AdS$ ($l=1$) spacetime, the increase in $\gamma$ increases $R_s$ whereas the opposite is observed in case of the black holes in Minkowski spacetime ($l\to \infty$). The charge $Q$ also has an effect on $R_s$. We observe that the shadow radius $R_s$ decreases with increase in charge $Q$ in both $AdS$ and asymptotically flat (Minkowski) spacetime. We then introduce a plasma background in order to observe the effect of refractive index ($n$) of the medium on the unstable circular photon orbits. It is observed that an increase in the plasma parameter results in decrease in the radius of the shadow $R_s$. The effect of the refractive index on the silhouette of the black hole shadow is similar in both spacetimes (asymptotically $AdS$ and Minkowski). Finally we compute the energy emission rate of the charged Gauss-Bonnet black hole and represent them graphically. We observe that the energy emission rate decreases with increase in the value of $\gamma$ in both $AdS$ and asymptotically flat spacetimes. We would like to mention that one of the main motivations of this work is to probe the signature of higher curvature correction in the shadow of black holes which in turn also brings out possibilities of looking for extra dimensions. In future we would like to investigate the effect of the spin parameter on Gauss-Bonnet black hole shadows.\\
	
	\section*{Acknowledgements}
	A.D. would like to acknowledge the support of S.N. Bose National Centre for Basic Sciences for Junior Research Fellowship. A.S. acknowledges the support by Council of Scientific and Industrial Research (CSIR, Govt. of India) for Junior Research Fellowship. S.G. acknowledges the support of the Visiting Associateship programme of IUCAA, Pune.

\end{document}